\documentclass[useAMS,usenatbib,levelone]{mn2e}
\usepackage{local_commands}
\usepackage{aas_macros}
\usepackage{amsmath}
\usepackage{amssymb}
\usepackage[dvips]{graphicx}
\usepackage{subfigure}
\title[Formation of a disc gap induced by a planet]{Formation of a disc gap induced by a planet: Effect of the deviation from Keplerian disc rotation}
\author[K. D. Kanagawa]{K.D. Kanagawa$^{1}$\thanks{E-mail:kanagawa@lowtem.hokudai.ac.jp}, H. Tanaka$^{1}$, T. Muto$^{2}$, T. Tanigawa$^{3}$ and T. Takeuchi$^{4}$ \\
$^{1}$Institute of Low Temperature Science, Hokkaido University, Sapporo 060-0819, Japan\\
$^{2}$Division of Liberal Arts, Kogakuin University, 1-24-2, Nishi-Shinjuku, Shinjuku-ku, Tokyo, 163-8677, Japan\\
$^{3}$School of Medicine, University of Occupational and Environmental Health, Yahatanishi-ku, Kitakyushu, Fukuoka 807-8555, Japan \\
$^{4}$Department of Earth and Planetary Sciences, Tokyo Institute of Technology, Meguro-ku, Tokyo 152-8551, Japan}
\begin{document}

\date{\today}

\pagerange{\pageref{firstpage}--\pageref{lastpage}} \pubyear{2014}

\maketitle
\label{firstpage}

\begin{abstract}
	The gap formation induced by a giant planet is important in the evolution of the planet and the protoplanetary disc.
	We examine the gap formation by a planet with a new formulation of one-dimensional viscous discs which takes into account the deviation from Keplerian disc rotation due to the steep gradient of the surface density.
	This formulation enables us to naturally include the Rayleigh stable condition for the disc rotation.
	It is found that the derivation from Keplerian disc rotation promotes the radial angular momentum transfer and makes the gap shallower than in the Keplerian case.
	For deep gaps, this shallowing effect becomes significant due to the Rayleigh condition.
	In our model, we also take into account the propagation of the density waves excited by the planet, which widens the range of the angular momentum deposition to the disc.
	The effect of the wave propagation makes the gap wider and shallower than the case with instantaneous wave damping.
	With these shallowing effects, our one-dimensional gap model is consistent with the recent hydrodynamic simulations.
\end{abstract}
\begin{keywords}
	accretion, accretion discs, protoplanetary discs, planets and satellites: formation
\end{keywords}

\section{Introduction} \label{sec:intro}
% what is a disk gap
A planet in a protoplanetary disc gravitationally interacts with the disc and exerts a torque on it.
The torque exerted by the planet dispels the surrounding gas and forms a disc gap along the orbit of the planet \citep{Lin_Papaloizou1979,Goldreich_Tremaine1980}.
However, a gas flow into the gap is also caused by viscous diffusion and hence the gap depth is determined by the balance between the planetary torque and the viscous diffusion.
Accordingly, only a large planet can create a deep gap \citep{Lin_Papaloizou1993,Takeuchi_Miyama_Lin1996,Ward1997,Rafikov2002,Crida_Morbidelli_Masset2006}.

% importance of a disk gap
The gap formation strongly influences the evolution of both the planet and the protoplanetary disc in various ways.
For example, a deep gap prevents disc gas from accreting onto the planet and slows down the planet growth \citep{D'Angelo_Henning_Kley2002,Bate_Lubow_Ogilvie_Miller2003,Tanigawa_Ikoma2007}, and also changes the planetary migration from the type I to the slower type II \citep{Lin_Papaloizou1986b,Ward1997}.
Furthermore, a sufficiently deep gap inhibits gas flow across the gap \citep{Artymowicz_Lubow1996,Lubow_Seibert_Artymowicz1999,Kley1999}, which is a possible mechanism for forming an inner hole in the disc \citep{Zhu_etal2011,Dodson-Robinson_Salyk2011}.

% discrepancy between 1d model and hydrodynamic simulation
Because of their importance, disc gaps induced by planets have been studied by many authors, using simple one-dimensional disc models \citep[e.g.,][]{Takeuchi_Miyama_Lin1996,Ward1997,Crida_Morbidelli_Masset2006,Lubow_D'Angelo2006} and numerical hydrodynamic simulations \citep{Artymowicz_Lubow1994,Kley1999,Varnire_Quillen_Frank2004,Duffell_MacFadyen2013,Fung_etal2013}.
One-dimensional disc models predict an exponential dependence of the gap depth.
That is, the minimum surface density at the gap bottom is proportional to $\exp[-A(\qplanet/\qstar)^{2}]$, where $\qplanet$ and $\qstar$ are the masses of the planet and the central star, and $A$ is a non-dimensional parameter (see also eq.~[\ref{eq:smin_kepler}]).
On the other hand, recent high-resolution hydrodynamic simulations done by \citeauthor{Duffell_MacFadyen2013} (\citeyear{Duffell_MacFadyen2013}, hereafter DM13) show that the gap is much shallower for a massive planet than the prediction of one-dimensional models.
According to their results, the minimum surface density at the gap is proportional to $(\qplanet/\qstar)^{-2}$.
\cite{Varnire_Quillen_Frank2004} and \cite{Fung_etal2013} obtained similar results from their hydrodynamic simulations.
Its origin has not yet been clarified by the one-dimensional disc model.
\cite{Fung_etal2013} also estimated the gap depth with a ``zero-dimensional'' analytic model, by simply assuming that the planetary gravitational torque is produced only at the gap bottom.
Their simple model succeeds in explaining the dependence of the minimum surface density of $\propto (\qplanet/\qstar)^{-2}$.
However, the zero-dimensional model does not give the radial profile of the surface density (or the width of the gap).
It is not well understood what kind of profile accepts their assumption on the planetary torque.
Further development of the one-dimensional gap model is required in order to clarify both the gap depth and width.
Such a model enables us to connect the gaps observed in protoplanetary discs with the embedded planets.
%However, its origin has not yet been clarified.
%\cite{Fung_etal2013} also estimated the gap depth with a simple analytic model called the ``zero-dimensional model'' and succeeded in explaining the dependence of the minimum surface density of $\propto (\qplanet/\qstar)^{-2}$.
%However, it does not give a profiles of surface density around a planet.
%Thus, further development of the one-dimensional model is required.
%Moreover, developing the one-dimensional model could possibly connect to explain a width of gaps observed in protoplanetary discs.

% deviation of disk rotation
One of the problems of the one-dimensional disc model is the assumption of the Keplerian rotational speed.
The disc rotation deviates from the Keplerian speed due to a radial pressure gradient \citep{Adachi_Hayashi_Nakazawa1976}.
When a planet creates a deep gap, the steep surface density gradient increases the deviation of the disc rotation significantly, which affects the angular momentum transfer at the gap (see Sections \ref{sub:orderestimate} and \ref{sub:basic_eq}).
Furthermore, a large deviation of the disc rotation can also violate the Rayleigh stable condition for rotating discs \citep{Chandrasekhar1939}.
A violation of the Rayleigh condition promotes the angular momentum transfer and makes the surface density gradient shallower so that the Rayleigh condition is only marginally satisfied \citep{Tanigawa_Ikoma2007,Yang_Menou2010}.
To examine such feedback on the surface density gradient, we should naturally include the deviation from the Keplerian disc rotation in the one-dimensional disc model.

%% density wave propagation
Another simplification is in the wave propagation at the disc--planet interaction.
The density waves excited by planets radially propagate in the disc and the angular momenta of the waves are deposited on the disc by damping.
This angular momentum deposition is the direct cause of the gap formation.
Most previous studies simply assume instantaneous damping of the density waves after their excitation \citep[e.g.,][]{Ward1997,Crida_Morbidelli_Masset2006}.
If the wave propagation is taken into account, the angular momentum is deposited in a wider region of the disc, which increases the width of the gap \citep{Takeuchi_Miyama_Lin1996,Rafikov2002}.
In a wide gap, the disc--planet interaction would be weak because the disc gas around the planet decreases over a wide region.
Hence, we cannot neglect the effect of wave propagation on the gap formation.

% construction
In the present paper, we re-examine the gap formation by a planet with the one-dimensional disc model, taking into account the deviation from Keplerian rotation and the effect of wave propagation.
To include the deviation from Keplerian disc rotation, we modify the basic equations for one-dimensional accretion discs, detailed in the next section.
The effect of the wave propagation is included using a simple model.
In Section~\ref{sec:gapdepth_simple}, we obtain estimates of gap depths for two simple cases.
One estimate for a wide gap corresponds to the zero-dimensional model proposed by \cite{Fung_etal2013}.
In Sections~\ref{sec:gap_nominal} and \ref{sec:gap_tqvariants}, we present numerical solutions of the gap without and with wave propagation, respectively.
We find that the gap becomes shallow due to the effects of the deviation from Keplerian rotation, the violation of the Rayleigh condition and the wave propagation.
With these shallowing effects, our results are consistent with the recent hydrodynamic simulations.
In Section~\ref{sec:conclusions}, we summarize and discuss our results.
\section{Model and basic equations} \label{sec:basic_equations_disk_with_planet}
We examine an axisymmetric gap in the disc surface density around a planet by using the one-dimensional model of viscous accretion discs.
Although the Keplerian angular velocity is assumed in most previous studies, we take into account a deviation from Keplerian disc rotation in our one-dimensional model.
The deviation cannot be neglected for a deep gap, as will shown below.
We also assume non-self-gravitating and geometrical thin discs.
For simplicity, the planet is assumed to be in a circular orbit.
We also adopt simple models for density wave excitation and damping to describe the gap formation.

\subsection{Angular velocity of a protoplanetary disc with a gap} \label{sub:orderestimate}
The angular velocity, $\Omega$, of a gaseous disc around a central star with mass $\qstar$ is determined by the balance of radial forces:
\begin{align}
	\Omega^2 R - \frac{G\qstar}{R^2} - \frac{1}{\rhosurf}\pdev{P_{\rm 2D}}{R} &=0 \label{eq:radial_force_balance},
\end{align}
where $R$ is the radial distance from the central star, $\rhosurf$ is the surface density of the disc and $P_{\rm 2D}$ denotes the vertically averaged pressure.
On the left-hand side of equation~(\ref{eq:radial_force_balance}), the first term represents the centrifugal force on a unit mass of the disc.
The second and third terms are the gravitational force by the central star and the force of the radial pressure gradient, respectively.
For $P_{\rm 2D}$, we adopt the simple equation of state $P_{\rm 2D}=\sonic^2 \rhosurf$, where $\sonic$ is the isothermal sound speed.
Using this equation of state, equation~(\ref{eq:radial_force_balance}) can be rewritten as
\begin{align}
	\Omega^2&=\Omegak^2\left( 1- 2 \eta \right) \label{eq:omega},
\end{align}
with
\begin{align}
	\eta &= -\frac{\hp^2}{2R} \left( \pdev{\ln \rhosurf}{R} + \pdev{\ln \sonic^2}{R} \right) \label{eq:eta},
\end{align}
where $\Omegak=\sqrt{G\qstar/R^3}$ is the Keplerian angular velocity, and $\hp=\sonic/\Omegak$

For a disc with no gap, the order of magnitude of the non-dimensional parameter $\eta$ is $\mathrm{O}(\hp^2/R^2)$ \citep{Adachi_Hayashi_Nakazawa1976}, because the term in parentheses in equation~(\ref{eq:eta}) is comparable to $\sim 1/R$.
On the other hand, if a planet opens a deep gap with a width of $\sim \hp$, the steep gradient of the surface density increases $\eta$ to $\mathrm{O}(\hp/R)$.
Hence, it also enhances the deviation of the disc rotation from the Kepler rotation in the deep gap.
We neglect the term $\partial c/\partial R$ in equation~(\ref{eq:eta}) because the temperature gradient would be small.
Neglecting the smaller terms of $\mathrm{O}(\hp/R)$, we approximately obtain $\partial \Omega/\partial R$ as
\begin{align}
	\pdev{\Omega}{R}&= -\frac{3\Omegak}{2R}\left[ 1-\frac{\hp^2}{3} \pddev{\ln \rhosurf}{R} \right] \label{eq:domega}.
\end{align}
Note that the second term in the parentheses is of order unity since $d^2 \ln \rhosurf/dx^2 \sim 1/\hp^2$ in a deep gap.
Therefore, it is found that $\partial \Omega/\partial R$ is significantly altered from the Keplerian value due to the steep gradient of the surface density, though the deviation of $\Omega$ is small ($\sim \hp/R$).
As shown later, this deviation promotes radial viscous transfer of angular momentum and makes a gap shallower.

\subsection{Basic equations describing a disc gap around a planet} \label{sub:basic_eq}
The equations for conservation of mass and angular momentum are given by
\begin{align}
	\pdev{\rhosurf}{t}+\frac{1}{2\pi R}\pdev{\mflux}{R}&=\macc \label{eq:mass_transfer},
\end{align}
and 
\begin{align}
	\pdev{}{t}\left( \rhosurf \junit \right) + \frac{1}{2\pi R}\pdev{\jflux}{R} &= \junit \macc + \frac{1}{2\pi R} \Lambdep \label{eq:j_transfer},
\end{align}
where $\mflux$ and $\jflux$ are the radial fluxes of mass and angular momentum, and $\junit(=R^2\Omega)$ is the specific angular momentum.
In equation~(\ref{eq:mass_transfer}), the source term $\macc$ represents the mass accretion rate onto a unit surface area of the disc.
The accretion of disc gas onto the planet can be included in $\macc$ as a negative term.
In equation~(\ref{eq:j_transfer}), $\Lambdep(R)$ represents the deposition rate of the angular momentum from the planet on the ring  region with radius $R$.

To describe the deposition rate $\Lambdep$, we consider the angular momentum transfer from the planet to the disc.
This transfer process can be divided into two steps.
First the planet excites a density wave by the gravitational interaction with the disc \citep[e.g.,][]{Goldreich_Tremaine1980}.
Second, the density waves are gradually damped due to the disc viscosity or a nonlinear effect \citep{Takeuchi_Miyama_Lin1996,Goodman_Rafikov2001}.
As a result of the wave damping, the angular momenta of the waves are deposited on the disc.
If instantaneous wave damping is assumed, the deposition rate $\Lambdep$ is determined only by the wave excitation.
In Section~\ref{sec:wave_excitation}, we will describe the deposition rate for the case with instantaneous wave damping.
In Section~\ref{sec:wave_damping}, we will give a simple model of $\Lambdep$ for the case of gradual wave damping.

The radial angular momentum flux $\jflux$ is given by \citep[e.g.,][]{Lynden-Bell_Pringle1974}
\begin{align}
	\jflux&=\junit \mflux - 2\pi R^3 \nu \rhosurf \pdev{\Omega}{R} \label{eq:jflux_inf}.
\end{align}
The first term is the advection transport by the disc radial mass flow, $\mflux$, and the second term represents the viscous transport.
For the kinetic viscosity, we adopt the $\alpha$ prescription, i.e., $\nu = \alpha \sonic \hp$ \citep{Shakura_Sunyaev1973}.
Note that $\jflux$ does not include the angular momentum transport by the density waves in our formulation.

Equations~(\ref{eq:mass_transfer})--(\ref{eq:jflux_inf}) describe the time evolution of the three variables $\rhosurf$, $\mflux$ and $\jflux$ with the given mass source term $\macc$, the angular momentum deposition rate from a planet $\Lambdep$ and the disc angular velocity $\Omega$.
Note that $\Omega$ depends on $\partial \rhosurf/\partial R$, as in equations~(\ref{eq:omega}) and (\ref{eq:eta}).

Next we consider the disc gap in a steady state $(\partial/\partial t =0)$.
The time scale for the formation of a steady gap is approximately equal to the diffusion time within the gap width, $t_{\rm diff}=\hp^2/\nu$.
For a nominal value of $\alpha$ ($\sim 10^{-3}$), the diffusion time is roughly given by $10^{3}$ Keplerian periods, which is shorter than the growth time of planets ($10^{5\mbox{--}7}$ yr) \citep{Kokubo_Ida2000,Kokubo_Ida2002} or the life time of protoplanetary discs ($10^{6\mbox{--}7}$ yr) \citep{Haisch_Lada_Lada2001}.
Hence the assumption of a steady gap would be valid.
In addition, we assume $\macc=0$ for simplicity.
Although gas accretion onto the planet occurs for $\qplanet \gtrsim 10 \mear$ \citep{Mizuno1980,Kanagawa_Fujimoto2013}, the assumption of $\macc=0$ would be valid if the accretion rate onto the planet is smaller than the radial disc accretion rate, $\mflux$.

Under these assumptions, equation (\ref{eq:mass_transfer}) shows that $\mflux$ is constant.
Equation~(\ref{eq:j_transfer}) yields
\begin{align}
	\jflux &= \jfluxinf - \int^{\infty}_{R} \Lambdep dR' \label{eq:jflux},
\end{align}
where $\jfluxinf$ is the angular momentum flux without the planet.
From equations~(\ref{eq:jflux_inf}) and (\ref{eq:jflux}), we obtain
\begin{align}
	\junit \mflux - 2\pi R^3 \rhosurf \nu \dpar{\Omega}{R} &= \jfluxinf-\int^{\infty}_{R} \Lambdep dR' \label{eq:steady_disk}.
\end{align}
Equation (\ref{eq:steady_disk}) with a constant mass accretion rate describes a steady disc gap around a planet for a given $\Lambdep$.
Since $d\Omega/dR$ is given by equation~(\ref{eq:domega}), equation~(\ref{eq:steady_disk}) is the second-order differential-integral equation.
Note that equation~(\ref{eq:steady_disk}) is derived from equation~(\ref{eq:j_transfer}) and indicates the angular momentum conservation.

By differentiating equation~(\ref{eq:steady_disk}), we obtain a rather familiar expression for the mass flux:
\begin{align}
	\mflux  &=\left( \dpar{\junit}{R}  \right)^{-1} \left[ \dpar{}{R}\left( 2\pi \rp^3 \nu \rhosurf \dpar{\Omega}{R} \right) + \Lambdep \right] \label{eq:balance_tqdens}.
\end{align}
Note that this expression is valid only for the steady state.
In a time-dependent case, equation~(\ref{eq:balance_tqdens}) should include the term $- 2\pi R \rhosurf (\partial j/\partial t)$ in the parentheses.
As a boundary condition, the disc surface density should approach its unperturbed values at both sides of the gap far from the planet.

Here, we also consider the unperturbed surface density.
In the unperturbed state, $\Omega$ can be replaced by $\Omegak$, by neglecting the smaller term $\mathrm{O}(\hp^2/R^2)$ (see equations~[\ref{eq:omega}] and [\ref{eq:eta}]).
Furthermore, setting $\Lambdep=0$ in equation~(\ref{eq:steady_disk}), we obtain the unperturbed surface density $\rhosurf_0$ as
\begin{align}
	3\pi R^2 \nu \Omegak \rhosurf_0(R)&=-\junitk \mflux + \jfluxinf \label{eq:rhosurf0}.
\end{align}
Thus $\rhosurf_0$ is given by
\begin{align}
	\rhosurf_0&=-\frac{\mflux}{3\pi \nu}\left( 1-\frac{\jfluxinf}{R^2 \Omegak \mflux} \right).
	\label{eq:rhosurf02}
\end{align}
This agrees with the well-known solution for steady viscous accretion discs \cite[e.g.,][]{Lynden-Bell_Pringle1974}.

\subsection{Rayleigh condition} \label{sub:rayleigh}
For a deep gap around a large planet, the derivative of the angular velocity deviates significantly  from the Keplerian velocity, as shown in Section \ref{sub:orderestimate}.
A sufficiently large deviation in $\Omega$ violates the so-called Rayleigh stable condition of $d\junit/dR\ge0$ \citep[see,][]{Chandrasekhar1961}.
Such a steep gap is dynamically unstable, which would cause a strong angular momentum transfer, lessening the steepness of the gap.
This would make the unstable region marginally stable (i.e.,$d\junit/dR=0$).

Using equation~(\ref{eq:domega}), we give $d\junit/dR$ as
\begin{align}
	\frac{d\junit}{dR}&= \frac{1}{2} \rp \Omegakp \left( 1+\hpp^2 \ddpar{\ln \rhosurf}{R} \right),
	\label{eq:djdr}
\end{align}
where the suffix $\planet$ indicates the value at $R=\rp$; this suffix is also used for other quantities.
Hence, using the second-derivative of the surface density, the marginally stable condition $d\junit/dR=0$ can be rewritten as 
\citep{Tanigawa_Ikoma2007}.
\begin{align}
	\hpp^2 \ddpar{\ln \rhosurf}{R} &=-1.
	\label{eq:marginal_condition}
\end{align}

Actually, around a sufficiently large planet, equation~(\ref{eq:steady_disk}) gives $\hpp^2 d^2\ln \rhosurf/d R^2 < -1$ in some radial regions.
In such unstable regions, we have to use equation~(\ref{eq:marginal_condition}) instead of equation~(\ref{eq:steady_disk}).

The breakdown of equation~(\ref{eq:steady_disk}) indicates that the flux $\jflux$ of equation~(\ref{eq:jflux_inf}) cannot transport all of the angular momentum deposited by the planet.
In a real system, however, the instability would enhance the angular momentum flux, which keeps the gap marginally stable.
The enhancement of $\jflux$ can be considered to be due to an effective viscosity $\nueff$ enhanced by the instability.
Since such an effective viscosity restores equation~(\ref{eq:steady_disk}), $\nueff$ in the unstable region is given by
\begin{align}
	\nueff&={\displaystyle  \frac{- \junit \mflux + \jfluxinf - \int^{\infty}_{R} \Lambdep dR'}{4\pi R^2 \rhosurf \Omega}} \label{eq:nu_eff},
\end{align}
where we use the relation $d\Omega/dR=-2\Omega/R$ obtained from the marginally stable condition.
Furthermore, by using $\nueff$ instead of $\nu$, equation~(\ref{eq:jflux_inf}) gives the enhanced angular momentum flux in the unstable region.

% 1/6 revised
The Rossby wave instability may be important for the gap formation \citep[e.g.,][]{Richard_Barge_LeDizes2013,Zhu_Stone_Rafikov2013,Lin2014}.
%The Rossby wave instability relates to the disc rotation as the same as the Rayleigh condition \citep{Li_etal2000}.
As well as the Rayleigh condition, the Rossby wave instability relates to the disc rotation \citep{Li_etal2000}.
Because it can occur before the Rayleigh condition is violated, however, the Rossby wave instability may suppress the surface density gradient more than the Rayleigh condition.
%The Rossby wave instability may suppress the surface density gradient more than the Rayleigh condition, though this instability relates to the disc rotation as the same as the Rayleigh condition \citep{Li_etal2000}.
For simplicity, we include only the Rayleigh condition in the present study.
A further detail treatment including the Rossby wave instability should be done in future works.

%\subsection{Excitation of density wave} \label{sec:wave_excitation}
\subsection{Angular momentum deposition from a planet} \label{sec:amd}
In the disc--planet interaction, a planet excites density waves and the angular momenta of the waves are deposited on the disc through their damping.
The angular momentum deposition rate $\Lambdep$ is determined by the later process.
Firstly, we will consider the deposition rate $\Lambdep$ in the case with instantaneous wave damping.
In this case, the deposition rate is governed only by the wave excitation.
Next, taking into account the wave propagation before damping, we will model the deposition rate in a simple form.

\subsubsection{Case with instantaneous wave damping} \label{sec:wave_excitation}
Under the assumption of instantaneous wave damping, the angular momentum deposition rate $\Lambdep(R)$ is equal to the excitation torque density $\Lambex(R)$, which is the rate at which a planet adds angular momenta to density waves per unit radial distance at $R$.
That is,
\begin{align}
	\Lambdep &= \Lambex.
	\label{eq:lambda_insitu}
\end{align}
At a position far from the planet, the excitation torque density is given by the WKB formula \citep[e.g.,][]{Ward1986} as 
\begin{align}
	\Lambex^{\rm WKB}=&{\displaystyle \pm C \pi \rp^2 \rhosurf \left( \frac{\qplanet}{\qstar} \right)^{2} \left(\rp \Omegakp \right)^2 \left( \frac{\rp}{R-\rp} \right)^{4}},
	\label{eq:wkb_tqdens}
\end{align}
where $C=(2^5/3^4)[2K_0(2/3)+K_1(2/3)]^2/\pi \simeq 0.798$ and $K_i$ denote the modified Bessel functions.
The sign of equation~(\ref{eq:wkb_tqdens}) is positive for $R>\rp$ or negative for $R\leq \rp$.
In the close vicinity of the planet, $|R-\rp| \lesssim \hpp$, on the other hand, the WKB formula is overestimated.
Thus, we model the excitation torque density $\Lambex$ with a simple cutoff as 
\begin{align}
	\Lambex&=
	\begin{cases}
		\Lambex^{\rm WKB} &\;\for \ \ \  |R-\rp| > \hpp \Delta,\\
		0                 &\;\for \ \ \  |R-\rp| \leq \hpp \Delta.
	\end{cases}
	\label{eq:tqdens}
\end{align}
The cut-off length $\hpp \Delta$ is determined so that the one-sided torque $T$ $(=\int^{\infty}_{\rp} \Lambda_{\rm ex} dR$) agrees with the result of the linear theory for realistic discs \citep{Takeuchi_Miyama1998,Tanaka_Takeuchi_Ward2002,Muto_Inutsuka2009}.
Then we obtain $\Delta=1.3$.

Note that the WKB formula is derived for discs with no gap.
\cite{Petrovich_Rafikov2012} reported that the torque density is altered by the steep gradient of the surface density because of the shift of the Lindblad resonances.
For simplicity, however, we ignore this effect in the present paper.
Hence, in our model, the excitation torque density $\Lambex$ is simply proportional to the disc surface density at $R$, $\rhosurf(R)$, and is independent of the surface density gradient even for deep gaps.
For a large planet with a mass of $\qplanet/\qstar \gtrsim (\hpp/\rp)^3$; furthermore, the non-linear effect would not be negligible for wave excitation \citep{Ward1997,Miyoshi1999}.
This non-linear effect is also neglected in our simple model.

\subsubsection{Case with wave propagation} \label{sec:wave_damping}
When wave propagation is included, the angular momentum deposition occurs at a different site from the wave excitation and equation~(\ref{eq:tqdens}) is not valid.
In this case, the angular momentum deposition is also governed by the damping of the waves.
Although the wave damping has been examined in previous studies \citep[e.g.,][]{Takeuchi_Miyama_Lin1996,Korycansky_Papaloizou1996,Goodman_Rafikov2001}, it is not clear yet how the density waves are damped in a disc with deep gaps.
In the present study, therefore, we adopt a simple model of angular momentum deposition, described below.

Since the waves are eventually damped in the disc, the one-sided torque (i.e., the total angular momentum of the waves excited at the outer disc in unit time) is equal to the total deposition rate in the steady state.
That is,
\begin{align}
	\toneside&=\int^{\infty}_{\rp} \Lambex dR' = \int^{\infty}_{\rp} \Lambdep dR'.
	\label{eq:def_onesided_torque}
\end{align}
Using the one-sided torque, the angular momentum deposition rate can be expressed by 
\begin{align}
	\Lambdep&=\pm \toneside f(R) \label{eq:lambda_propagation},
\end{align}
where the distribution function $f(R)$ satisfies $\int^{\infty}_{\rp} f(R)dR=1$, and the sign is the same as in equation~(\ref{eq:wkb_tqdens}).
As a simple model, we assume a distribution function $f(R)$ given by 
\begin{align}
	f(R)&=
	\begin{cases}
		{\displaystyle \frac{1}{\wdep}} &\ \for\ \ \ {\xd\hpp-\frac{\wdep}{2}<|R-\rp|<\xd\hpp+\frac{\wdep}{2}},\\
		0     &\;\mbox{otherwise}.\\
	\end{cases}
	\label{eq:fx_rectangle}
\end{align}
In this simple model, the non-dimensional parameter $\xd$ determines the position of the angular momentum deposition and the parameter $\wdep$ represents the radial width of the deposition site.
The waves propagate from the excitation site to the deposition site around $|x|=\xd$.
Since the density waves propagate away from the planet, the deposition site is farther from the planet than the excitation site.
The parameter $\xd$ should be consistent with this condition.

In the case with wave propagation, we use equations~(\ref{eq:lambda_propagation}) and (\ref{eq:fx_rectangle}) to obtain the gap structure with equation~(\ref{eq:steady_disk}).
It should be noted that $T$ in equation~(\ref{eq:lambda_propagation}) depends on the surface density distribution through the definition of equation~(\ref{eq:def_onesided_torque}), because $\Lambex$ is proportional to $\rhosurf$.
These coupled equations are solved as follows.
First, we obtain the surface density distribution with equation~(\ref{eq:steady_disk}) for a given $T$.
Next, we determine the corresponding mass of the planet from equation~(\ref{eq:def_onesided_torque}), using the obtained surface density.

\subsection{Local approximation and non-dimensional equations} \label{sec:nond_eq}
The typical width of a disc gap is comparable to the disc scale height and much smaller than the orbital radius of the planet.
Thus it is convenient to use the local coordinate defined by
\begin{align}
	x&=\frac{R-\rp}{\hpp} \label{eq:def_x}.
\end{align}
Note that the suffix $\planet$ indicates the value at $R=\rp$.

We adopt a local approximation in which terms proportional to $\hpp/\rp$ and higher order terms are neglected.
From equations~(\ref{eq:omega}) and (\ref{eq:eta}), the deviation in $\Omega$ from $\Omegak$ is given by
\begin{align}
	\Omega-\Omegak&= \frac{\hpp \Omegakp}{2\rp} \dpar{\ln \rhosurf}{x}, \label{eq:omega_local}
\end{align}
and is proportional to $\hpp/\rp$.
Thus, the disc angular velocity $\Omega$ is replaced by the angular velocity of the planet $\Omegakp$ under the local approximation, and the specific angular momentum $\junit$ also is given by $\rp^2\Omegakp$.
As for the derivative $d\Omega/dR$, we cannot neglect the deviation from the Keplerian value.
Equation~(\ref{eq:domega}) yields
\begin{align}
	\dpar{\Omega}{R}&=-\frac{3\Omegakp}{2\rp} \left( 1-\frac{1}{3}\ddpar{\ln \rhosurf}{x} \right) \label{eq:domega_local}.
\end{align}

Equation~(\ref{eq:steady_disk}) can be rewritten in the local approximation as
\begin{align}
	\junitkp \mflux + &3\pi \rp^2 \nup \rhosurf \Omegakp \left( 1-\frac{1}{3}\ddpar{\ln \rhosurf}{x} \right)\nonumber \\
	&=\jfluxinf-\int^{\infty}_{x} \Lambdep \hpp dx'.
	\label{eq:steady_disk2}
\end{align}
Because of the local approximation, equation~(\ref{eq:steady_disk2}) cannot be applied for the wide gap formation.
If the half width of gap is narrower than about $1/3 \rp$, equation~(\ref{eq:steady_disk2}) would be valid.

Here we introduce the non-dimensional surface density, $s$, defined by
\begin{align}
	s=\frac{\rhosurf}{\rhosurf_0(\rp)},\label{eq:s}
\end{align}
where $\rhosurf_0(\rp)$ is the unperturbed surface density at $R=\rp$ given by equation~(\ref{eq:rhosurf02}).
Dividing equation~(\ref{eq:steady_disk2}) by $3\pi \rp^2 \nup \rhosurf_0(\rp) \Omegakp$ and using equation~(\ref{eq:rhosurf0}), we obtain a non-dimensional form:
\begin{align}
	\left( 1-\frac{1}{3} \ddpar{\ln s}{x} \right)s &= 1- \frac{1}{3} \int^{\infty}_{x} \tLambda dx' \label{eq:sdif_zero_order2},
\end{align}
where $\tLambda$ is the non-dimensional angular momentum deposition rate defined by
\begin{align}
	\tLambda&=\frac{\Lambdep \hpp}{\pi \rp^2 \nup \rhosurf_0(\rp) \Omegakp}.\label{eq:nond_lambda}
\end{align}
The marginally stable condition can be rewritten as
\begin{align}
	\ddpar{\ln s}{x}&=-1 \label{eq:gap_breaking_rayleigh}.
\end{align}
This equation is used instead of equation~(\ref{eq:sdif_zero_order2}) in the Rayleigh unstable region.

The non-dimensional excitation torque density, $\tLambdaex$, is defined by
\begin{align}
	\tLambdaex=\frac{\Lambex \hpp}{\pi \rp^2 \nup \rhosurf_0(\rp) \Omegakp}&=
	\begin{cases}
		{\displaystyle \pm K \frac{ C}{x^{4}} } s(x) &\for \ \ \ |x|>\Delta,\\
		0                                            &\for \ \ \ |x|\leq \Delta.\\
	\end{cases}
	\label{eq:nond_tqdens}
\end{align}
where the non-dimensional parameter $K$ is given by
\begin{align}
	K&=\left( \frac{\qplanet}{\qstar} \right)^2 \left( \frac{\rp}{\hpp} \right)^{5} \alpha^{-1} \label{eq:def_k}.
\end{align}
In the above, we use $\nup=\alpha \hpp^2 \Omegakp$.
In our model, the parameter $K$ is the only parameter that determines the gap structure for the instantaneous damping case.

In the case with instantaneous wave damping, the angular momentum deposition rate is given by $\tLambda=\tLambdaex$ (eq.[\ref{eq:lambda_insitu}]).
In the case with wave propagation, equation~(\ref{eq:lambda_propagation}) gives
\begin{align}
	\tLambda(x)&=\ttilde \hpp f(\rp+\hpp x)
	\label{eq:nond_lambda_propagation},
\end{align}
where the non-dimensional one-sided torque, $\ttilde$, is given by
\begin{align}
	\ttilde&=K \int^{\infty}_{\Delta} \frac{C}{x^4} s(x) dx \label{eq:relate_ttilde_K}.
\end{align}
Note that the deposition rate $\tLambda$ includes two parameters $\xd$ and $\wdep$, in addition to $K$.
In this case, we solve equation~(\ref{eq:sdif_zero_order2}) for a given value of $\ttilde$.
Then, we can obtain $K$ by substituting the solution $s(x)$ into equation~(\ref{eq:relate_ttilde_K}), as mentioned at the end of the last subsection.
In order to obtain the solution for a certain $K$, we need an iteration of the above procedure with trial values of $\ttilde$.

The boundary conditions of equations~($\ref{eq:sdif_zero_order2}$) and (\ref{eq:gap_breaking_rayleigh}) are
\begin{align}
	s&=1, \; \mbox{ at  } x=\pm \infty.
	\label{eq:bc}
\end{align}
Under the local approximation, the surface density has a symmetry of $s(x)=s(-x)$, since both the above basic equations and the deposition rate are symmetric.

% ˆÚ"®'³'¹'é?
%Note that the gap becomes eccentric if the gap is extended to the 1:2 Lindblad resonance \citep{Kley_Dirksen2006,Fung_etal2013}.
%Because of the local approximation, however, such wide gaps are beyond the scope of this work.

\section{Estimates of gap depths for simple situations} \label{sec:gapdepth_simple}
\subsection{Case of the Keplerian discs}
Before deriving the gap solution in our model described in Section~\ref{sec:basic_equations_disk_with_planet}, we examine the gaps for two simple situations.
First, we consider a disc with Keplerian rotation, as assumed in previous studies. 
Neglecting the deviation in $d\Omega/dR$ from the Keplerian (i.e., the term of $d^2\ln s/dx^2$) in equation~(\ref{eq:sdif_zero_order2}), we have
\begin{align}
	s&=1-\frac{1}{3} \int^{\infty}_{x} \tLambda dx'
	\label{eq:skep}
\end{align}
Here we also assume instantaneous wave damping and adopt $\tLambda=\tLambdaex$ (eq.~[\ref{eq:nond_tqdens}]).
%The case with wave propagation will be discussed later.
Differentiating equation~(\ref{eq:skep}), we obtain
\begin{align}
	\dpar{\ln s}{x}&=
	\begin{cases}
		{\displaystyle \pm \frac{C}{3x^4}K } &\for \ \ \ |x|>\Delta,\\
		0               &\for \ \ \ |x|\leq \Delta,
	\end{cases}
	\label{eq:sdif_kepler}
\end{align}
Hence, we obtain the surface density in the Keplerian discs with the instantaneous wave damping as
\begin{align}
	s(x)&= 
	\begin{cases}
		{\displaystyle \exp\left( - \frac{C}{9|x|^3}    K \right) }&\for \ \ \ |x|>\Delta,\smallskip \\
		{\displaystyle \exp\left( - \frac{C}{9\Delta^3} K \right) }&\for \ \ \ |x| \leq \Delta,\\
	\end{cases}
	\label{eq:s_kepler}
\end{align}
Using equation~(\ref{eq:def_k}), $C=0.798$ and $\Delta=1.3$, the minimum surface density, $\smin$, is
\begin{align}
	\smin&= \exp \left[ -0.040 \alpha^{-1} \left( \frac{\rp}{\hpp} \right)^5 \left( \frac{\qplanet}{\qstar} \right)^2 \right] \label{eq:smin_kepler}.
\end{align}
This solution is almost the same as that in the previous one-dimensional gap model \citep[e.g.,][]{Lubow_D'Angelo2006}.

%When the Rayleigh criterion is violated in the case with a large $K$, equations~(\ref{eq:s_kepler}) and (\ref{eq:smin_kepler}) are no longer valid.
%\cite{Tanigawa_Ikoma2007} (TI07 in the following) obtain the gap structure which the Rayleigh condition is considered, which is described in Appendix~\ref{sec:tanigawa}.
%In addition, we can also construct the Keplerian solution in the case with wave propagation, which described in Appendix~\ref{sec:kep_propagation}.
For a very large $K$, the Rayleigh condition is violated and equations~(\ref{eq:s_kepler}) and (\ref{eq:smin_kepler}) are invalid.
\cite{Tanigawa_Ikoma2007} obtained the gap structure in Keplerian discs, including the Rayleigh condition.
Their solution is described in Appendix~\ref{sec:tanigawa}.
In Appendix~\ref{sec:kep_propagation}. we also derive gap solutions in Keplerian discs, taking into account the wave propagation with the simple model of equation~(\ref{eq:lambda_propagation}) and (\ref{eq:fx_rectangle}).

\subsection{Case of the wide-limit gap} \label{sec:widegap}
%Next, we consider a situation implied by the ``zero-dimension'' analysis done by \cite{Fung_etal2013}, which the gap is very wide.
%We call this situation 'wide-limit gap' case.
%In the case of the wide-limit gap, it is assumed that the density waves are excited only at the gap bottom with $s\simeq \smin$.
%Then, the one-sided torque of equation~(\ref{eq:relate_ttilde_K}) is simply given by
Next, we consider a situation implied by the zero-dimension analysis done by \cite{Fung_etal2013}, which assumes that the wave excitation occurs only a the gap bottom.
This assumption would be valid if the gap bottom region is wide enough.
Hence, we call this situation 'wide-limit gap' case.
Since the density waves are excited at the gap bottom with $s\simeq \smin$, the one-sided torque of equation~(\ref{eq:relate_ttilde_K}) is simply given by
\begin{align}
	\ttilde=\frac{C}{3\Delta^3}K \smin \simeq 0.121K\smin. \label{eq:ttilde_widegap}
\end{align}
Using equation~(\ref{eq:sdif_zero_order2}), we can estimate $\smin$ of the wide-limit gap.
The right-hand side of equation~(\ref{eq:sdif_zero_order2}) can be rewritten as $1-\ttilde/3$ at $x=0$.
In the left-hand side of equation~(\ref{eq:sdif_zero_order2}), moreover, we can neglect the term $d^2 \ln s/dx^2$ when a flat-bottom gap is assumed.
Then, the relation between $\smin$ and $\ttilde$ is obtained as
\begin{align}
	\smin &= 1-\frac{\ttilde}{3}. \label{eq:relation_smin_ttilde_widegap}
\end{align}
Equations~(\ref{eq:ttilde_widegap}) and (\ref{eq:relation_smin_ttilde_widegap}) yield
\begin{align}
	\smin=\frac{1}{1+0.040K} \label{eq:smin_widegap}.
\end{align}
For a large $K$, $\smin$ given by equation~(\ref{eq:smin_widegap}) is proportional to $1/K$.
This result agrees with the zero-dimensional model by \cite{Fung_etal2013}\footnote{In the notation of \cite{Fung_etal2013}, $K$ is given by $q^2/(\alpha [h/r]^5)$}.
In the zero-dimensional model, the minimum surface density is estimated from a balance between the planetary torque and the viscous angular momentum flux outside the gap.
Such a balance is also seen from equation~(\ref{eq:relation_smin_ttilde_widegap}) (and eq.~[\ref{eq:sdif_zero_order2}]).
The first and second terms in the right-hand side of equation~(\ref{eq:relation_smin_ttilde_widegap}) correspond to the viscous angular momentum flux outside the gap and the planetary torque and the left-hand side is negligibly small for a large $K$.

With their hydrodynamic simulations for $K \lesssim 10^{4}$, DM13 derived a similar result\footnote{In the notation of DM13, $K$ is given by ${\cal M}^{-1}(M_{\rm sh}/M_{\rm p})^{2} \alpha^{-1}$.}, 
\begin{align}
	\smin=\frac{29}{29+K} = \frac{1}{1+0.034K} \label{eq:smin_dm13}.
\end{align}
It is found that equations~(\ref{eq:smin_widegap}) and (\ref{eq:smin_dm13}) are consistent with each other.
%The wide-limit gap requires that all the waves are excited in the bottom region with $s\simeq \smin$.
%Therefore we should check whether or not the waves are excited in the bottom region in our solutions.
%We also note that the gap depth given by equation~(\ref{eq:smin_widegap}) is much shallower than that in the case of the Keplerian disc (eq.~[\ref{eq:smin_kepler}]) for a deep gap with a large $K$.
Note that these minimum surface densities are much larger than that of the Keplerian disc (eq.~[\ref{eq:smin_kepler}]) for a large $K$ because equation~(\ref{eq:ttilde_widegap}) is not accepted in the Keplerian solution.
The wide-limit gaps assume that all the waves are excited in the bottom region with $s\simeq \smin$, i.e., equation~(\ref{eq:ttilde_widegap}).
In Sections~\ref{sec:gap_nominal} and \ref{sec:gap_tqvariants}, we will check whether or not this assumption is valid, by comparing it with our one-dimensional solutions.

\section{Gap structure in the case with instantaneous wave damping} \label{sec:gap_nominal}
\subsection{Linear solutions for shallow gaps} \label{sub:shallow_gap}
Here we present the numerical solution of the gap in the case with instantaneous wave damping (i.e., $\tLambda=\tLambdaex$).

% linear solution
First, we consider the case with a small $K$ in equation~(\ref{eq:sdif_zero_order2}), in which $\tLambda$ is proportional to $K$.
This case corresponds to a shallow gap around a small planet.
Since $|s-1|$ is small, it is useful to express the solution as
\begin{align}
	s&=\exp(K y) \label{eq:y},
\end{align}
or $s=1+K y$.
As seen in the next subsection, the former expression is better for an intermediate $K$ ($\sim 10$).
Substituting equation~(\ref{eq:y}) into equation~(\ref{eq:sdif_zero_order2}) with equation~(\ref{eq:lambda_insitu}), we can expand it into a power series of $K y$.
The first order terms give the linear equation of $y$:
\begin{align}
	\ddpar{y}{x}-3y&=\mp
	\begin{cases}
		{\displaystyle\frac{C}{3 |x|^3}     }  &\for \ \ \ |x| >   \Delta,\smallskip \\
		{\displaystyle \frac{C}{3 \Delta^{3}}} &\mbox{otherwise},\\
	\end{cases}
	\label{eq:linear_eq}
\end{align}
where the sign in the right-hand side is negative for $x>0$ and positive for $x\leq 0$.
Equation~(\ref{eq:linear_eq}) is an inhomogeneous linear differential equation, and can be integrated with the boundary conditions of equation~(\ref{eq:bc}).
We do not need to take care of the Rayleigh condition in the shallow gaps.
A detailed derivation of the linear solution is described in Appendix~\ref{sec:particular_solution}.

\begin{figure}
	\begin{center}
		\resizebox{0.47\textwidth}{!}{\includegraphics{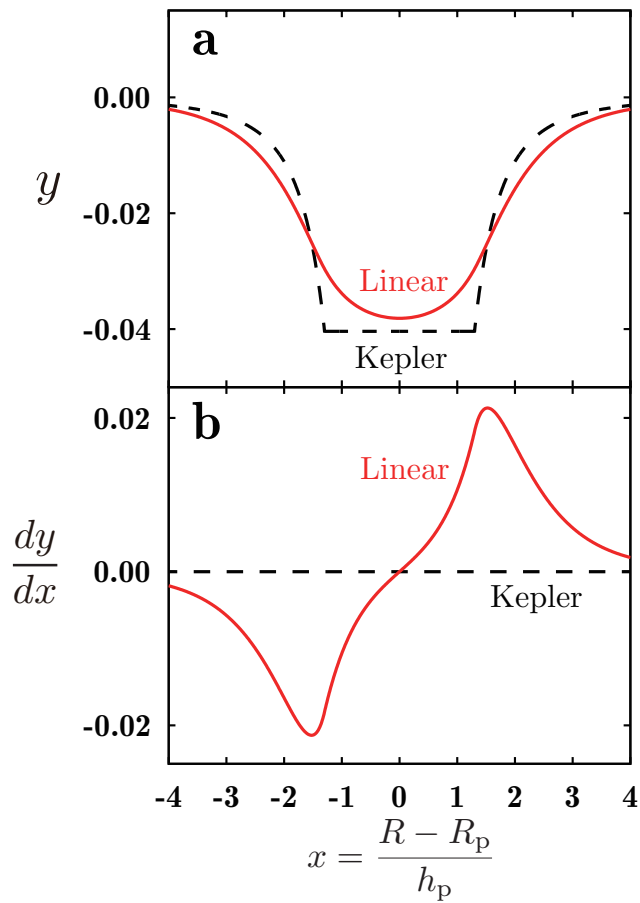}}
		\caption{
		Linear solution for $y$ (a) and $dy/dx$ (b).
		The surface density and angular velocity are given by $y$ and $dy/dx$ with equations~(\ref{eq:y}) and (\ref{eq:delta_omega}), respectively.
		The dashed line is the solution for the Keplerian disc.
		}
		\label{fig:lineargap_profile}
	\end{center}
\end{figure}
Fig.~\ref{fig:lineargap_profile}a shows $y$, which can be converted into the surface density $s$ by equation~(\ref{eq:y}).
In these shallow gaps, the gap depth is almost the same as for the Keplerian case, though our model gives a smooth surface density distribution.

Fig.~\ref{fig:lineargap_profile}b shows the derivative of $y$ which is related to $\Delta \Omega$, as
\begin{align}
	\Delta \Omega &= \Omega-\Omegak = K \frac{\hpp \Omegakp}{2\rp} \dpar{y}{x} \label{eq:delta_omega},
\end{align}
using equations~(\ref{eq:omega_local}) and (\ref{eq:y}).
The absolute value of $\Delta \Omega$ attains a maximum at $|x|\simeq 1.5$.
The second-order derivative of $y$ gives the shear, $d\Omega/dx$, as
\begin{align}
	\dpar{\Omega}{x}&=\dpar{\Omegak}{x} \left( 1-\frac{K}{3}\ddpar{y}{x} \right) \label{eq:shear_y},
\end{align}
as seen from equation~(\ref{eq:domega_local}).
At $|x|>1.5$, the shear motion is enhanced compared to the Keplerian case, because $d^2 y/dx^2 <0$.
Since the shear motion causes viscous angular momentum transfer, this enhancement makes the surface density gradient less steep compared with the Keplerian case, as shown in Fig.~\ref{fig:lineargap_profile}a.

\subsection{Nonlinear solutions for deep gaps}  \label{sub:gapprofile_standard}
Next we consider deep gaps around relatively large planets.
In this case, we numerically solve the non-linear equation~(\ref{eq:sdif_zero_order2}) with the Rayleigh condition.
We call the obtained non-linear solution the ``exact'' solution.

At regions far from the planet, the surface density perturbation is rather small and the linear approximation is valid.
Thus, we adopt a linear solution at $|x|>10$.
Note that this linear solution has different coefficients for the homogeneous terms from those in Section~\ref{sub:shallow_gap}(see Appendix~\ref{sec:particular_solution}).
The coefficients of the homogeneous solution are given to satisfy the boundary conditions of equation~(\ref{eq:bc}).
At $|x|\le 10$, we integrate equation~(\ref{eq:sdif_zero_order2}) with the fourth-order Runge-Kutta integrator.
In the Rayleigh unstable region, the surface density is governed by the marginally stable condition (eq.~[\ref{eq:gap_breaking_rayleigh}]), instead of equation~(\ref{eq:sdif_zero_order2}).

\begin{figure}
	\centering
	\resizebox{0.47\textwidth}{!}{\includegraphics{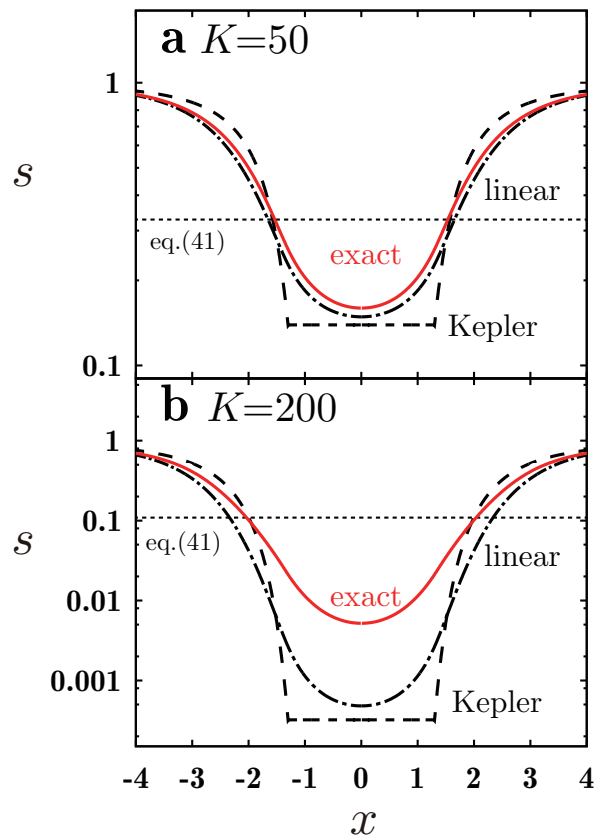}}
	% (a) K=50 (Rayleigh stable case)
	% (b) K=200 (Rayleigh unstable case)
	\caption{
	Surface density distributions for $K=50$ (a) and $200$ (b).
	The red line is the exact solution (see text).
	The chain line is the linear solution given by equation~(\ref{eq:y}) and the dashed line is the solution for the Keplerian case (eq.~[\ref{eq:s_kepler}]).
	The dotted line represents the minimum surface densities for the wide-limit gap given by equation~(\ref{eq:smin_widegap}).
	}
	\label{fig:shallow_and_deep_gaps}
\end{figure}
Fig.~\ref{fig:shallow_and_deep_gaps} shows the surface density distributions of the exact solutions for $K=50$ (a) and $200$ (b).
If we assume a disc with $\hpp/\rp=0.05$ and $\alpha=10^{-3}$, these cases correspond to $\qplanet=1/8\mjup$ and $1/4\mjup$ respectively, where $\mjup$ is the mass of Jupiter.
For comparisons, the Keplerian solution (eq.~[\ref{eq:s_kepler}]) and the linear solution with equation~(\ref{eq:y}) are also plotted.
For $K=50$, the linear solution almost agrees with the exact solution, while it is much deeper than the exact solution for $K=200$.
For $K=200$, the Keplerian solution has a much smaller $\smin$ than the exact solution.

\begin{figure}
	\centering
	\resizebox{0.47\textwidth}{!}{\includegraphics{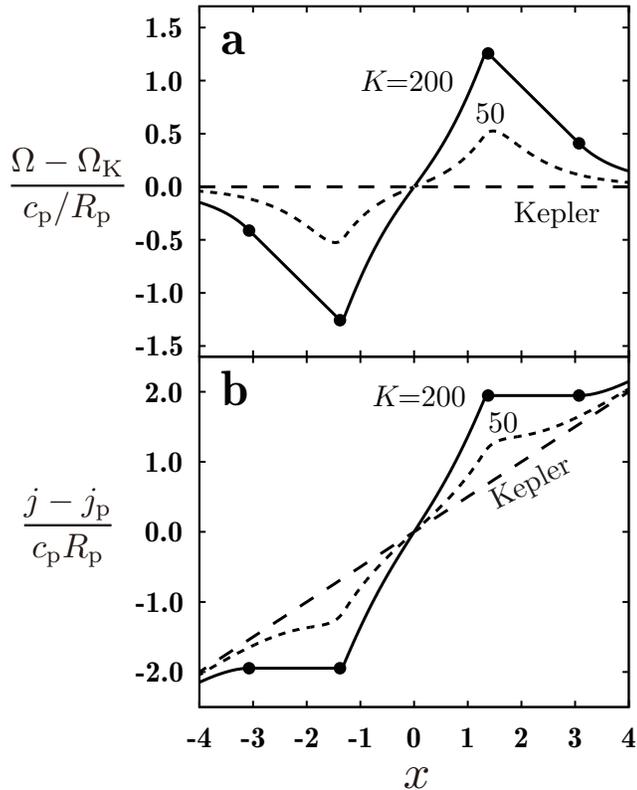}}
	% a angular velocity
	% b specific angular momentum
	\caption{
(a) Deviation from Keplerian disc rotation and (b) specific angular momentum, for $K=50$ (dashed) and $200$ (solid).
	The filled circles indicate the edge of the marginally stable region for the Rayleigh condition.
	}
	\label{fig:discrotation_instant}
\end{figure}
\begin{figure}
	\centering
	\resizebox{0.47\textwidth}{!}{\includegraphics{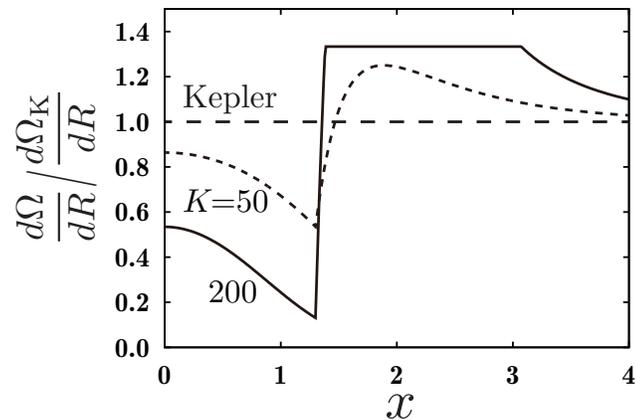}}
	\caption{
	Shear of exact solutions for $K=50$ (dashed) and $200$ (solid).
	}
	\label{fig:domega_instant}
\end{figure}
Fig.~\ref{fig:discrotation_instant} illustrates the angular velocities (a) and specific angular momenta (b) for the exact solutions for $K=50$ and $200$.
Similar to the linear solution in Fig.~\ref{fig:lineargap_profile}, the shear motion is enhanced at $|x| \gtrsim 1.4$.
This enhancement of the shear motion is also seen in Fig.~\ref{fig:domega_instant}.
The enhancement promotes the angular momentum transfer and makes the surface density gradient less steep.
For $K=200$, the Rayleigh condition is violated.
In the unstable region, the marginally stable condition further reduces the surface density gradient.
This makes the gap much shallower than for the Keplerian solution, as seen in Fig.~\ref{fig:shallow_and_deep_gaps}b.

\begin{figure}
	\centering
	\resizebox{0.47\textwidth}{!}{\includegraphics{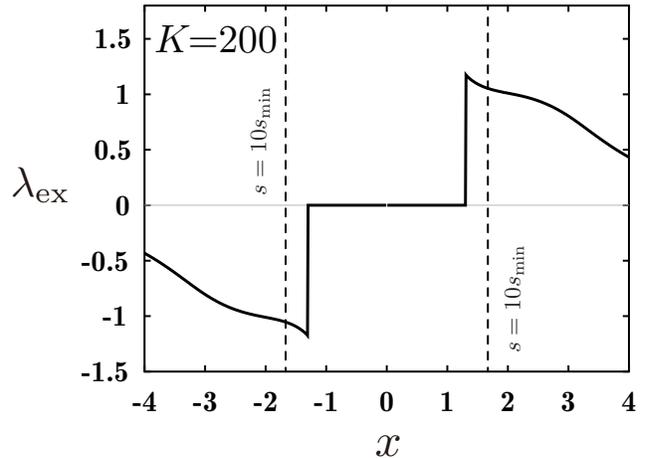}}
	\caption{
	Excitation torque density given by equation~(\ref{eq:nond_tqdens}) for $K=200$.
	The two vertical lines indicate the positions with $s=10\smin$.
	}
	\label{fig:tqdens_instant_K200}
\end{figure}
We also plot the minimum surface densities, $\smin$, of the wide-limit gap (eq.~[\ref{eq:smin_widegap}]) in Fig.~\ref{fig:shallow_and_deep_gaps}.
The wide-limit gap gives a much larger $\smin$ than the exact solution for $K=200$.
In the wide-limit gap, it is assumed that the density waves are excited only at the gap bottom with $s\simeq \smin$.
Fig.~\ref{fig:tqdens_instant_K200} shows the excitation torque density given by equation~(\ref{eq:tqdens}) for the exact solution with $K=200$.
This torque density indicates that the waves are excited mainly in the region with $s>10\smin$.
Thus the assumption of wave excitation at the gap bottom is not valid in this case.
Since wave excitation with a larger $s$ increases the one-sided torque, this can explain why the gap of the exact solution is much deeper than the wide-limit gap in Fig.~\ref{fig:shallow_and_deep_gaps}.
Note that this result for the wave excitation is obtained in the case of instantaneous wave damping.
The effect of the wave propagation can change the gap width and the mode of wave excitation, as seen in the next section.

\subsection{Effect of the Rayleigh condition} \label{sub:effect_rayleigh}
We further examine the effect of the Rayleigh condition on the gap structure.
\begin{figure}
	\centering
	\resizebox{0.47\textwidth}{!}{\includegraphics{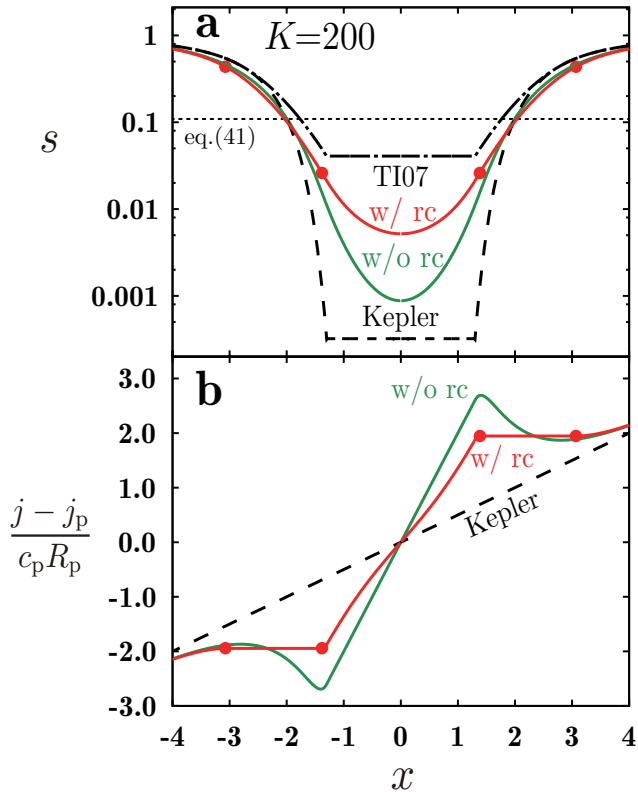}}
	\caption{
	(a) Surface density distribution and (b) specific angular momentum distribution, for $K=200$.
	The red line indicates the exact solution.
	The green line is the solution without the Rayleigh condition (see text).
	The chain line in (a) denotes the surface density distribution given by the model of \protect \cite{Tanigawa_Ikoma2007} (eq.~[\ref{eq:rhosurf_nominal_ti}], TI07).
	}
	\label{fig:gap_rayleigh}
\end{figure}
Fig.~\ref{fig:gap_rayleigh} shows the surface densities (a) and specific angular momenta (b) for the exact solution and the solution without the Rayleigh condition.
The solution without the Rayleigh condition has unstable regions with $d\junit/dx<0$ (i.e., $1.4<|x|<3.1$).
This comparison between these two solutions directly shows how the Rayleigh condition changes the gap structure.
The Rayleigh condition increases $\smin$ by a factor 6 for $K=200$.
This is because the marginal condition of $d^2 \ln s/dx^2 \geq -1$ keeps the surface density gradient less steep and makes the gap shallow.

It can be considered that the marginally stable state is maintained by $\nueff$ of equation~(\ref{eq:nu_eff}).
The non-dimensional form of equation~(\ref{eq:nu_eff}) is given by
\begin{align}
	\frac{\nueff}{\nu}&={\displaystyle \frac{3- \int^{\infty}_{x} \tLambda dx'}{4s}} \label{eq:nond_nueff}.
\end{align}
\begin{figure}
	\centering
	\resizebox{0.47\textwidth}{!}{\includegraphics{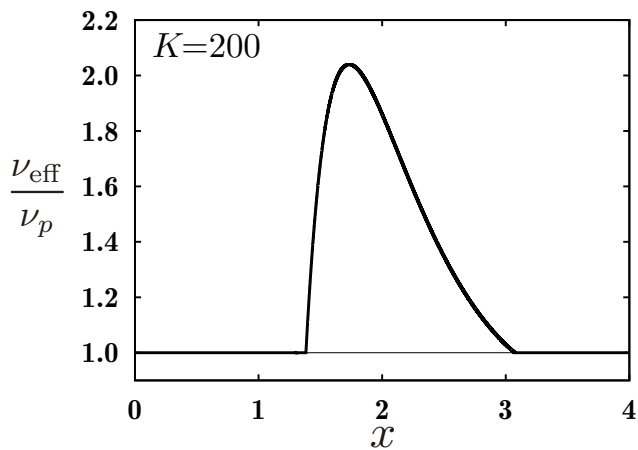}}
	\caption{
	Effective viscosity $\nueff$ of the exact solution with $K=200$.
	}
	\label{fig:nu_enhanced_rayleigh}
\end{figure}
Fig.~\ref{fig:nu_enhanced_rayleigh} shows $\nueff$ in the unstable region for $K=200$.
The effective viscosity is twice as large as the original value at $x=1.8$.
This enhancement of the effective viscosity causes the shallowing effect in Fig.~\ref{fig:gap_rayleigh}a.

%In Fig.~\ref{fig:gap_rayleigh}a, we also plot the surface density distribution given by the model of (TI07).
%TI07 also took into account the Rayleigh condition in their gap model, though they assumed Keplerian disc rotation (for details, see Appendix~\ref{sec:tanigawa}).
In Fig.~\ref{fig:gap_rayleigh}a, we also plot the surface density distribution given by \cite{Tanigawa_Ikoma2007}(hereafter TI07), in which the Rayleigh condition is taken into account (for details, see Appendix~\ref{sec:tanigawa}).
Their model gives a shallower gap than our exact solution.
This is because a very steep surface density gradient in the Keplerian solution is suppressed by the Rayleigh condition to a greater extent than in our model.

%We show that the Keplerian solution given by TI07 does not conserve the angular momentum.
%Since the Keplerian solution without the Rayleigh condition (eq.~[\ref{eq:s_kepler}]) comes from equation~(\ref{eq:skep})(or equation~[\ref{eq:sdif_zero_order2}]) which is originated by equation~(\ref{eq:steady_disk}), the solution satisfies the angular momentum conservation.
%When the Rayleigh criterion is violated, equation~(\ref{eq:s_kepler}) is no longer valid.
%TI07 also considered the Rayleigh condition by adopting the marginal condition (eq.~[\ref{eq:gap_breaking_rayleigh}]) instead of equation~(\ref{eq:s_kepler}) in an unstable region.
%%The Rayleigh condition is considered by adopting marginal condition (eq.~[\ref{eq:gap_breaking_rayleigh}]) instead of equation~(\ref{eq:s_kepler}) in an unstable region.
%The surface density at the flat-bottom of TI07 (eq.~[\ref{eq:smin_tanigawa}]) does not satisfy equation~(\ref{eq:skep}) and the angular momentum conservation.
%On the other hand, this breakdown is resolved in our new formulation.
%Our exact solutions in Fig.\ref{fig:gap_rayleigh} (red solid line) come from equation~(\ref{eq:sdif_zero_order2}) and satisfy the angular momentum conservation.
%Hence, we have to take into account the deviation from the Keplerian rotation to satisfy the angular momentum conservation.
We also show that the Keplerian solution by TI07 does not satisfy the angular momentum conservation.
The Keplerian solution without the Rayleigh condition~(eq.~[\ref{eq:s_kepler}]) is derived just from equation~(\ref{eq:skep}) (or eq.~[\ref{eq:sdif_zero_order2}]), which is originated from equation~(\ref{eq:steady_disk}).
In this solution, thus, the angular momentum conservation is satisfied.
However, when the Rayleigh condition is violated, the marginal stable condition (eq.~[\ref{eq:gap_breaking_rayleigh}]) is used instead of equation~(\ref{eq:s_kepler}).
Because of this, the surface density at the flat bottom of TI07's solution does not satisfy equation~(\ref{eq:skep}) or the angular momentum conservation, either.
This violation is resolved in our formulation because our exact solution always satisfies equation~(\ref{eq:sdif_zero_order2}) outside the Rayleigh unstable region.\footnote{By introducing the effective viscosity of equation~(\ref{eq:nond_nueff}) and multiplying the LHS of equation~(\ref{eq:sdif_zero_order2}) by $\nueff/\nu_0$, equation~(\ref{eq:sdif_zero_order2}) is recovered in the Rayleigh unstable region.}

\subsection{Gap depth} \label{sec:gapdepth_nominal}
\begin{figure}
	\centering
	\resizebox{0.47\textwidth}{!}{\includegraphics{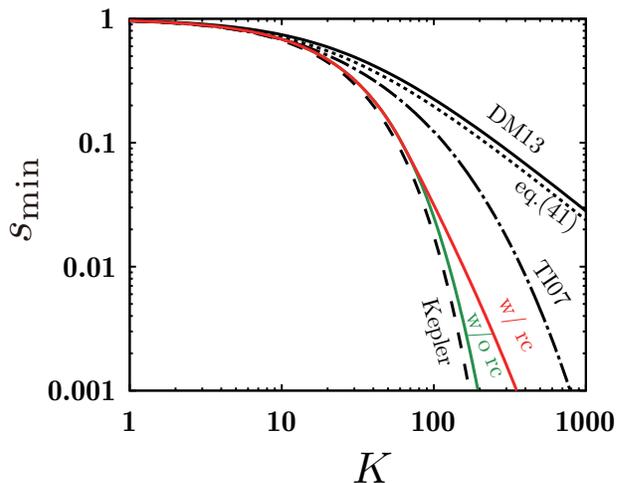}}
	\caption{
	Minimum surface densities, $\smin$, for the exact solution (red line) and the solution without the Rayleigh condition (green line).
	The dashed line is $\smin$ in the Keplerian case.
	The chain, dotted and solid lines denote $\smin$ given by the model of TI07, the wide-limit gap (eq.~[\ref{eq:smin_widegap}]) and the empirical relation of DM13 (eq.~[\ref{eq:smin_dm13}]) respectively.
	}
	\label{fig:depth_rayleigh}
\end{figure}
Fig.~\ref{fig:depth_rayleigh} shows the minimum surface densities, $\smin$, as a function of $K$ for the exact solutions.
For comparison, we also plot $\smin$ for the solutions without the Rayleigh condition and the Keplerian solutions.
These solutions give deeper gaps than the exact solution, similar to the result of Section~\ref{sub:gapprofile_standard} .
It is found that the shallowing effect due to the Rayleigh condition becomes significant with an increase in $K$.
This is because the Rayleigh condition is violated more strongly for large $K$.

In Fig.~\ref{fig:depth_rayleigh}, on the other hand, the exact solution is much deeper than DM13's results and the wide-limit gap, though the latter two cases agree well with each other.
The model of TI07 also gives much deeper gaps than DM13.
These comparisons indicate that in the case with instantaneous wave damping, our exact solution cannot reproduce the hydrodynamic simulations of DM13.
This difference in the gap depth from DM13 is likely to be due to the fact that the assumption of the wide-limit gap is not satisfied in the case with instantaneous wave damping (see Fig.~\ref{fig:tqdens_instant_K200}).
In the next section, we will see that the effect of wave propagation widens the gap and makes the assumption of the wide-limit gap valid.
\section{Effect of density wave propagation} \label{sec:gap_tqvariants}
In this section, we consider the effect of wave propagation.
Wave propagation changes the radial distribution of the angular momentum deposition.
A simple model of angular momentum deposition rate altered by wave propagation is described in Section~\ref{sec:wave_damping}.
Using this simple model, we solve equation~(\ref{eq:sdif_zero_order2}) with the Rayleigh condition in the similar way to the previous section.
At the region far from the planet (i.e., $|x|>10$), we use the linear solution to equation~(\ref{eq:dequation1}) with $g(x)=0$ in this case.

\subsection{Gap structure for $K=200$} \label{sub:gap_propagation}

\begin{figure}
	\centering
	\resizebox{0.47\textwidth}{!}{\includegraphics{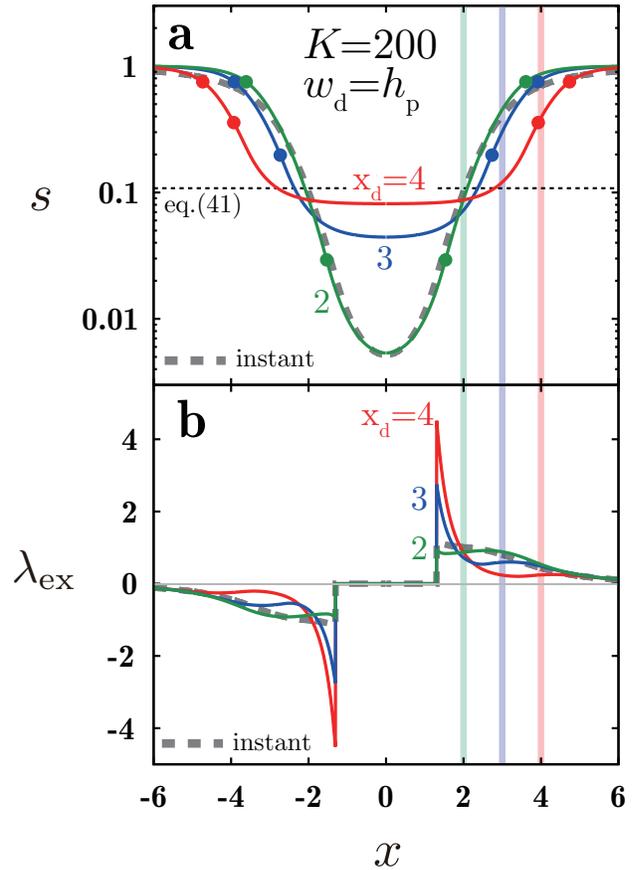}}
	\caption{
	Surface densities (a) and excitation torque densities (b) in the case with the wave propagation for $K=200$.
	The green, blue and red lines denote the solutions with $\xd=2$, $3$ and $4$ respectively.
	The parameter $\wdep$ is set to $\hpp$.
	The gray dashed line is the surface density in the case with instantaneous wave damping.
	The dotted line in (a) represents the minimum surface density for the wide-limit gap given by equation~(\ref{eq:smin_widegap}), i.e. $\smin=0.109$.
	}
	\label{fig:gap_rectangle_xdvar}
\end{figure}
Fig.~\ref{fig:gap_rectangle_xdvar} illustrates the surface densities (a) and the excitation torque densities (b) of the exact solutions in the case with wave propagation.
The angular momenta of the excited waves are deposited around $|x|=\xd$ in our model.
A large $\xd$ indicates a long propagation length between the excitation and the damping.
The parameter $K$ is set to $200$.
For an increasing $\xd$, the gap becomes wider and shallower.
The gap width is directly governed by the position of the angular momentum deposition.
For $\xd=3$ and $4$, the gap depths are consistent with the wide-limit gap (and also DM13).
For $\xd=4$, the density waves are excited mainly at the bottom region with $s\simeq \smin$, as seen in Fig.~\ref{fig:gap_rectangle_xdvar}b.
Moreover, for $\xd=3$, a major part of the wave excitation occurs at the bottom.
That is, the assumption of the wide-limit gap is almost satisfied for the solutions with $\xd=3$ and $4$.
This explains why the gap depths are consistent with the wide-limit gap for these large $\xd$.

It is also valuable to compare the gap width with hydrodynamic simulations.
DM13 performed a simulation for the case of $\qplanet=1/4\mjup$ ($2M_{\rm sh}$ in their notation), $\alpha=10^{-3}$ and $\hpp/\rp=0.05$.
This case corresponds to $K=200$.
In this simulation, they found that the gap width is about $6\hpp$, assuming that these gap edges are located at the position with $\rhosurf = (1/3) \rhosurf_0(\rp)$ (i.e., $s=1/3$).
If we adopt the same definition of the gap edge, the gap widths of our exact solutions with $\xd=3$ and $4$ are $6.1\hpp$ and $7.7\hpp$, respectively.
Hence, if we take into account the wave propagation and adopt $\xd=3$--$4$, our exact solution can almost reproduce both of the gap width and depth of the hydrodynamic simulations by DM13, for $K=200$.

It should be also noted that, for $\xd=2$, the wave excitation mainly occurs at $|x|>\xd$ ($80\%$ of the excitation torques come from this region).
However, the deposition site should be farther from the planet than the excitation site because the density waves propagate away from the planet.
Thus, the case with $\xd=2$ does not represent a realistic wave propagation.
From now on, we judge that our simple model for the wave propagation is valid if more than half of the one-sided torque arises from the excitation at $|x|<\xd$.
In the case with $\xd=3$ or $4$, the excitation at $|x|<\xd$ contributes $55\%$ or $78\%$ of the one-sided torque, respectively.

\begin{figure}
	\centering
	\resizebox{0.47\textwidth}{!}{\includegraphics{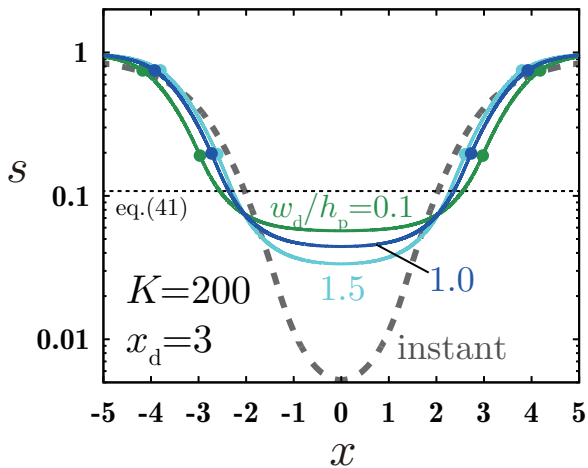}}
	\caption{
	Gap structures for $\wdep = 0.1\hpp$ (green), $\hpp$ (blue) and $1.5\hpp$ (light-blue).
	The parameters $K$ and $\xd$ are set to $200$ and $3$, respectively.
	The gray dashed line indicates the solution in the instantaneous damping case and the dotted line is the minimum surface density of the wide-limit gap (eq.~[\ref{eq:smin_widegap}]).
	}
	\label{fig:gap_rectangle_wdvar}
\end{figure}
In Fig.~\ref{fig:gap_rectangle_wdvar}, we check the effect of the width of the deposition site, $\wdep$, for $\xd=3$ and $K=200$.
It is found that the width $\wdep$ has only a small influence on the gap structure.

\begin{figure}
	\centering
	\resizebox{0.47\textwidth}{!}{\includegraphics{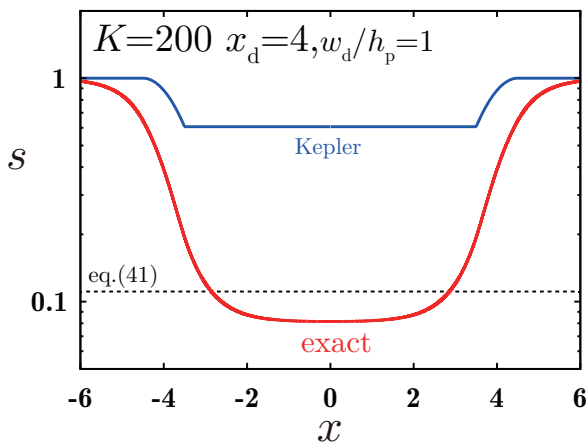}}
	\caption{
	The Keplerian solution in the case of wave propagation for $K=200$, $\xd=4$ and $\wdep=\hpp$ (bule line).
	For comparison, the exact solution (red line) is also plotted.
	}
	\label{fig:gap_rectangle_compKepler}
\end{figure}
We show that the deviation from the Keplerian rotation is also important in the case with wave propagation.
In Fig.~\ref{fig:gap_rectangle_compKepler}, we plot the solution with the Keplerian rotation and our exact solution.
The Keplerian solution is derived from equation~(\ref{eq:skep}) with the angular momentum deposition model (eqs.~[\ref{eq:lambda_propagation}] and [\ref{eq:fx_rectangle}]).
When the Rayleigh condition is violated, the marginal stable condition (eq.~[\ref{eq:gap_breaking_rayleigh}]) is used.
A detail derivation of this solution is described in Appendix~\ref{sec:kep_propagation}.
In the Keplerian solution of Fig.~\ref{fig:gap_rectangle_compKepler}, the Rayleigh condition is violated over the whole region of the angular momentum deposition.
Then the minimum surface density is given by equation~(\ref{eq:smin_kepler_propagation_wrc}), which is much larger than our solution and equation~(\ref{eq:smin_widegap}).
Because equation~(\ref{eq:smin_kepler_propagation_wrc}) does not satisfy equation~(\ref{eq:skep}), the Keplerian solution does not satisfy the angular momentum conservation, as pointed out in Section~\ref{sub:effect_rayleigh}.
On the other hand, in the zero-dimension analysis by \cite{Fung_etal2013} (or in eq.~[\ref{eq:smin_widegap}]), $\smin$ is estimated from a balance between the planetary torque and the viscous angular momentum flux (i.e., from the angular momentum conservation).
Because of this difference, the Keplerian solution gives a much shallower gap than the estimation in equation~(\ref{eq:smin_widegap}).
Note that because our exact solutions are given by equation~(\ref{eq:sdif_zero_order2}), the balance between the planetary torque and the viscous angular momentum flux is always satisfied in our solutions.
Hence, our soltuions always satisfy the angular momentum conservation and gives a similar $\smin$ to equation~(\ref{eq:smin_widegap}) for a sufficiently wide gap.

\subsection{Dependences of the gap depth and width on $K$}
\begin{figure}

	\centering
	\resizebox{0.47\textwidth}{!}{\includegraphics{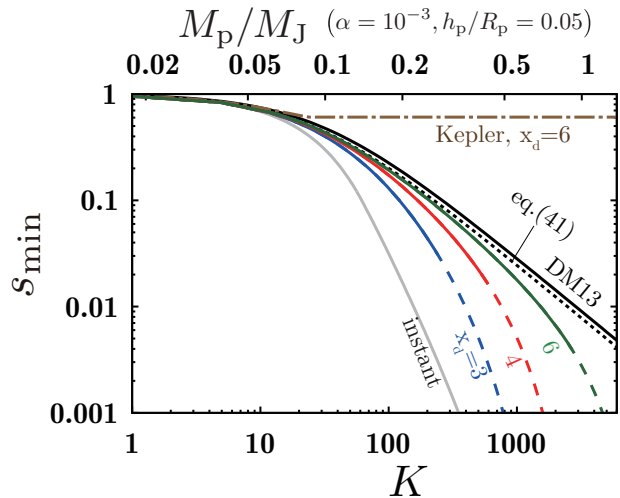}}
	\caption{
	Minimum surface densities, $\smin$, of the exact solutions in the case with the wave propagation for $\xd=3$ (blue), $4$ (red) and $6$ (green).
	The parameter $\wdep$ is $\hpp$.
	We also plot results by DM13 (eq.~[\ref{eq:smin_dm13}], solid line) and the wide-limit gap (eq.~[\ref{eq:smin_widegap}], dotted line) and the exact solution with instantaneous damping (gray line).
	The dashed lines indicate exact solutions with unrealistic wave propagation.
	}
	\label{fig:depth_gap_rectangle_xdvar}
\end{figure}
Fig.~\ref{fig:depth_gap_rectangle_xdvar} shows the minimum surface densities $\smin$ as a function of the parameter $K$ similar to Fig.~\ref{fig:depth_rayleigh} but the effect of the wave propagation is included in this figure.
In this figure, we also show the dependence on the parameter $\xd$ while $\wdep$ is fixed at $\hpp$ since $\wdep$ does not change the surface density distribution much (see Fig.~\ref{fig:gap_rectangle_wdvar}).
At $K=200$, as also seen in Fig.~\ref{fig:gap_rectangle_xdvar}, our exact solutions reproduce the gap depth of DM13 (or the wide-limit gap) for $\xd\geq 3$.
At $K=1000$, on the other hand, a larger $\xd$ ($\geq 6$) is required for agreement with DM13.
That is, with an increase of $K$, a large $\xd$ is necessary for the values of the depth of the hydrodynamic simulations to agree.
Note that the dashed lines in Fig.~\ref{fig:depth_gap_rectangle_xdvar} represent the cases of unrealistic wave propagation, in which more than half of the one-sided torque is due to the excitation at $|x|>\xd$, as for the case of $\xd=2$ in Fig.~\ref{fig:gap_rectangle_xdvar}.
At large $K$, a large $\xd$ is also required for realistic wave propagation.

We also show the Keplerian solution with the wave propagation of $\xd=6$ (see Appendix~\ref{sec:kep_propagation}).
For $K>30$, $\smin$ is given by equation~(\ref{eq:smin_kepler_propagation_wrc}) and independent of $K$ because of the Rayleigh condition, as seen in Fig.~\ref{fig:gap_rectangle_xdvar}.
This unrealistic result in the Keplerian solution is related with the violation of the angular momentum conservation, as pointed out in Section~\ref{sub:effect_rayleigh} (and see also Appendix~\ref{sec:kep_propagation}).

\begin{figure} 
	\centering
	\resizebox{0.47\textwidth}{!}{\includegraphics{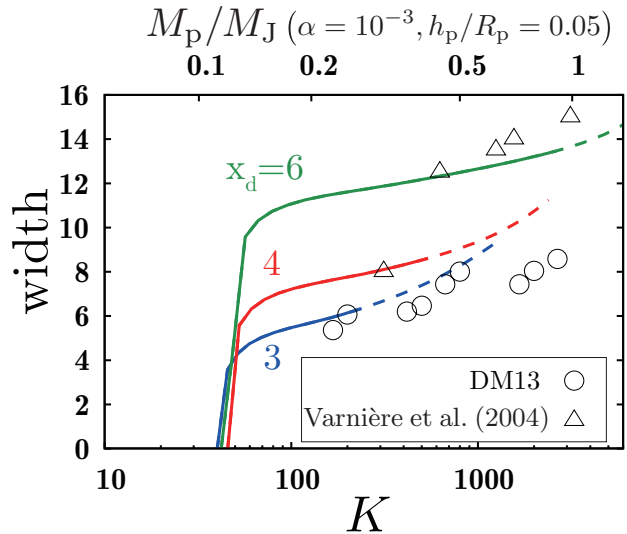}}
	\caption{Gap widths of our solutions for $\xd=3$ (blue), $4$ (red) and $6$ (green).
	The edge of the gap is defined by the position with $s=1/3$, in the same way as defined by DM13.
	The parameter $\wdep$ is set to $\hpp$.
	The dashed lines represent solutions with unrealistic wave propagation, similar to Fig.~\ref{fig:depth_gap_rectangle_xdvar}.
	The circles indicate the gap widths obtained by DM13 (Fig.~6 of their paper), and the triangles show the gap widths by \protect \cite{Varnire_Quillen_Frank2004} (twice $\Delta_{1000}$ in table~1 of \protect \citeauthor{Varnire_Quillen_Frank2004}).
	}
	\label{fig:width_gap_rectangle_xdvar}
\end{figure}
To check which $\xd$ is preferable, we also compare the gap width with the hydrodynamic simulations.
Fig.~\ref{fig:width_gap_rectangle_xdvar} shows the gap width of the exact solutions as a function of $K$.
Similar to DM13, the gap edge is defined by the position with $s=1/3$.
In this definition, the gap width is roughly given by twice $\xd \hp$ for our exact solutions with $K > 50$.
Note that this definition is useless for $K < 50$ because of shallow gaps with $\smin > 1/3$.
The dashed lines represent the cases of unrealistic wave propagation, similar to Fig.~\ref{fig:depth_gap_rectangle_xdvar}.
The results of DM13 and \cite{Varnire_Quillen_Frank2004} are also plotted in Fig.~\ref{fig:width_gap_rectangle_xdvar}.
\cite{Varnire_Quillen_Frank2004} also performed hydrodynamic simulations of gap formation for $\qplanet/\qstar = 10^{4}$ - $2\times 10^{3}$, $\alpha = 6 \times 10^{-2}$ - $6 \times 10^{-5}$ and $\hpp/\rp=0.04$ (i.e., $K=600$ - $6\times 10^{5}$).
Their gap depths almost agree with DM13's relation.
For $K<300$, our exact solutions with $\xd=3$ and $4$ agree with the results of DM13 and \cite{Varnire_Quillen_Frank2004}, respectively.
For $K>300$, on the other hand, the widths obtained by \cite{Varnire_Quillen_Frank2004} are wider than those given by DM13.
Our exact solution with $\xd=6$ agrees with the widths of \cite{Varnire_Quillen_Frank2004}, while widths of DM13 correspond to our solutions of unrealistic wave propagation.
The preferable $\xd$ cannot be determined only by this comparison, and we still have a large uncertainty in the preferable value of $\xd$.

The difference of widths between DM13 and \cite{Varnire_Quillen_Frank2004} would be caused by different parameters in their simulations (e.g., the disc viscosity, spatial resolution and width of a computation domain).
However, the origin of the difference is still unclear.
Note that the results of \cite{Kley_Dirksen2006} and \cite{Fung_etal2013} may support the wide gap formation of \cite{Varnire_Quillen_Frank2004}.
%According to \cite{Kley_Dirksen2006}, the gap becomes eccentric if the gap is extended to the 1:2 Lindblad resonance.
%The eccentric gaps are formed for a large $K$ ($\gtrsim 8000$) in the results of \cite{Kley_Dirksen2006} and \cite{Fung_etal2013}, which would indicate the wide gap formation, though such wide gaps are beyond the scope of present work because of the local approximation.
\cite{Kley_Dirksen2006} also showed that the disc rotation has some eccentricity when the gap is extended to the 1:2 Lindblad resonance.
The eccentric gaps are formed for a large $K$ ($\gtrsim 10^{4}$) \citep[e.g.,][]{Kley_Dirksen2006,Fung_etal2013}.
Such wide gaps by massive giant planets are beyond the scope of our one-dimensional disc model adopting the local approximation.

In the above, we found that a larger $\xd$ is required for a larger $K$ (i.e., a massive planet) in order to reproduce the minimum surface densities derived by DM13 and \cite{Varnire_Quillen_Frank2004}.
It should be noted that the propagation distance is not proportional to the parameter $\xd$.
The propagating distance of waves is defined by the distance from the wave excitation site to the angular momentum deposition site (i.e., $\xd$).
Since the one-sided toruqe is radially distributed (see Fig~\ref{fig:gap_rectangle_xdvar}b), the wave excitation site can be approximately given by the median of the distribution, i.e., the point within the half of the one-sided torque arises.
Such a excitation site shifts away from the planet with an increase of $K$ (see Fig.~\ref{fig:depth_gap_rectangle_xdvar})\footnote{
In Fig.~\ref{fig:depth_gap_rectangle_xdvar}, the exact solution have transition points from the realistic wave propagation (solid lines) to the unrealistic one (dashed lines) for each $\xd$.
At the transition point of $K$, the excitation site defined by the median is equation to $\xd$.
Fig.~\ref{fig:depth_gap_rectangle_xdvar} shows that the excitation site moves away from the planet with an increasing $K$ since the transition point of $K$ increases with $\xd$.}
.
% 12/26 revised
\cite{DAngelo_Lubow2010} also showed this tendency that the peak of the excitation torque density shifts away from the planet as a deep gap is formed (Fig.{15} in that paper), using hydrodynamic simulations.
Hence, because of this shift of the excitation site, the propagating distance does not increase much as $\xd$ with $K$.
\cite{Goodman_Rafikov2001} showed that the propagating distance of waves decreases with an increase of the planet mass due to the non-linear wave damping.
Hence, further studies are needed in order to confirm whether this results given by \cite{Goodman_Rafikov2001} conflicts with ours because of the shift of the excitation site.
Furthermore, the non-linear wave damping would be weakened by the steep surface density gradient at the gap edge, as pointed out by \cite{Petrovich_Rafikov2012}.
In order to fix the parameter $\xd$, such a wave damping effect in the gap should be taken into account in future work.

\section{Summary and Discussion} \label{sec:conclusions}
We re-examined the gap formation in viscous one-dimensional discs with a new formulation.
In our formulation, we took into account the deviation from Keplerian disc rotation and included the Rayleigh stable condition, consistently.
We also examined the effect of wave propagation.
Our results are summarized as follows.
\begin{enumerate}
	\item The deviation from the Keplerian disc rotation makes the gap shallow.
		This is because of the enhancement of the shear motion and the viscous angular momentum transfer at the gap edges (see Fig.~\ref{fig:domega_instant}).
	\item For deep gaps, the deviation from the Keplerian disc rotation is so large that the Rayleigh stable condition is violated.
		An enhanced viscosity dissolves such unstable rotation and makes it marginally stable (see Fig.~\ref{fig:nu_enhanced_rayleigh}).
		This effect also makes the gap shallower (see Fig.~\ref{fig:gap_rayleigh}).
	\item To include the effect of wave propagation, we adopted a simple model where the position of the angular momentum deposition is parameterized by $\xd$.
		A large $\xd$ indicates a long propagation length.
		The effect of wave propagation makes the gap wider and shallower (Fig.~\ref{fig:gap_rectangle_xdvar}).
		In a wide gap, the waves are mainly excited at the flat bottom, which reduces the one-sided torque and the gap depth.
		For a sufficiently large $\xd$, the gap depth of our exact solution agrees well with the wide-limit gap and with the results of hydrodynamic simulations.
		At $K=1000$, our model requires $\xd \geq 6$ for the agreement (Fig.~\ref{fig:depth_gap_rectangle_xdvar}).
		In the case of instantaneous wave damping, on the other hand, our exact solution gives much deeper gaps than those of hydrostatic simulations.
	\item To check the validity of the large $\xd$, the gap width of our exact solution is compared with results of hydrodynamic simulations.
		For $K=1000$, our exact solution with $\xd \geq 6$ has a gap width of $12\hpp$, which is larger than those of DM13 ($\sim 8\hpp$).
		The gap widths of \cite{Varnire_Quillen_Frank2004}, on the other hand, are almost consistent with our exact solutions.
		Because of this uncertainty in the gap width of hydrodynamic simulations, it is difficult to fix the preferable $\xd$ by this comparison.
	\item When the Rayleigh condition is taken into account, the deviation from the Keplerian rotation should also be included in order to keep the angular momentum conservation.
		The Keplerian solutions with the Rayleigh condition give much shallower gaps, as shown in figures~\ref{fig:depth_rayleigh} and \ref{fig:depth_gap_rectangle_xdvar}.
\end{enumerate}

In future works, we need to determine the preferable value of $\xd$.
Previous studies \citep[e.g.,][]{Takeuchi_Miyama_Lin1996,Korycansky_Papaloizou1996,Goodman_Rafikov2001,Dong_Rafikov_Stone2011} have investigated the wave propagation with no gap.
As pointed out by \cite{Petrovich_Rafikov2012}, however, the gap structure can affect the wave damping.
Since our result shows that the wave damping significantly affects both the gap depth and width, the wave damping should be treated accurately in both one-dimensional models and hydrodynamic simulations for gap formation.
%However, the gap formation would affect the wave damping, in addition to the wave excitation, as pointed out by \cite{Petrovich_Rafikov2012}.
%We should investigate the wave propagation and damping in deep gaps.
%Our result shows that the wave damping significantly affects both the gap depth and width.
%Hence, the wave damping should be treated accurately in hydrodynamic simulations of gap formation.

Our simple model does not include the effect of the deviation from Keplerian disc rotation on the wave excitation.
%\cite{Petrovich_Rafikov2012} showed that a steep surface density gradient modifies the wave excitation from that used in the present paper.
%We need to include this effect in further works.
\cite{Petrovich_Rafikov2012} showed that a steep surface density gradient modifies the excitation torque.
Such a effect on the wave excitation should be included in future studies on the gap formation.
Nevertheless, it is also considered that when the waves are mainly excited at the flat-bottom, such as for the wide-limit gap, the deviation of the disc rotation would not affect the wave excitation significantly.

We also neglect the non-linearity of wave excitation, whereas the non-linearity cannot be neglected for large planets as $\qplanet/\qstar \gtrsim (\hpp/\rp)^3$.
According to \cite{Miyoshi1999}, the non-linearity makes the excitation torque small compared to the value for linear theory.
This possibly leads to an additional shallowing effect.
However, this effect would not significantly influence the gap depth since $\smin$ is scaled by only $K$ in DM13's relation (eq.~[\ref{eq:smin_dm13}]).
%However, this effect may be a mirror since $\smin$ is scaled by only $K$ in DM13's relation (eq.~[\ref{eq:smin_dm13}]).

% 1/5 revised
%The Rossby wave instability may make gaps further shallower than our model since our model only includes the Rayleigh condition but not Rossby wave instability.
The Rossby wave instability may be essential for the gap formation. In present study, we included only the Rayleigh condition.
A more detail investigation including both the Rayleigh condition and the Rossby wave instability should be done in future works.

\section{acknowledgments}

We are grateful to Aur\'{e}lien Crida, and Alessandro Morbidelli for their valuable comments.
We also thank the anonymous referee for useful comments on the manuscript.
K.D.K. is supported by Grants-in-Aid for Scientific Research (26103701) from the MEXT of Japan.
T.T. is supported by Grants-in-Aid for Scientific Research (23740326 and 24103503) from the MEXT of Japan.

\appendix
\section{Gap model by Tanigawa\&Ikoma(2007)} \label{sec:tanigawa}
Following \cite{Tanigawa_Ikoma2007}, we describe surface density structures of gap in a disc with the Keplerian rotation.
In this section, we consider the gap structure in the case of instantaneous wave damping.
The gap structure in the case with wave propagation is discussed in Appendix~\ref{sec:kep_propagation}.
In the case of instantaneous wave damping, the angular momentum deposition rate is given by $\tLambdaex$ given by equation~(\ref{eq:nond_tqdens}).
At far from a planet, the Rayleigh condition is always satisfied because the planetary gravity is weak.
Hence, the gradient of surface density is given by equation~(\ref{eq:sdif_kepler}) and the second derivative of surface density is given by
\begin{align}
	\ddpar{\ln s}{x}&=\mp\frac{4KC}{3x^5} \label{eq:sdif2_ti}.
\end{align}
The stability of the Rayleigh condition is checked by equation~(\ref{eq:sdif2_ti}).
Namely, when $d^2\ln s/dx^2 < -1$ in equation~(\ref{eq:sdif2_ti}), the surface density is described by the marginal Rayleigh stable state, instead of equation~(\ref{eq:s_kepler}).
We should point out that the second derivative given by equation~(\ref{eq:sdif2_ti}) is used to determine the stability of the Rayleigh condition and does not affect the surface density distribution.
Because of this, the surface density gradient is steeper and the shallowing effect of the Rayleigh condition is much higher than in our model.
Using equation~(\ref{eq:nond_tqdens}), we obtain the outer edge of the marginal Rayleigh stable region, $x=x_m$, as
\begin{align}
	x_m&=\left( \frac{4}{3}CK \right)^{1/5} \label{eq:xm}.
\end{align}
In consideration of the continuity of the surface density distribution, the surface density in the marginal Rayleigh stable region ($x < x_m$) is given by
\begin{align}
	\ln s &=-\frac{5}{6}x_m^2+\frac{5}{4}x_m |x| -\frac{1}{2} x^2 \label{eq:rhosurf_nominal_ti},\\
	&=-0.854K^{2/5}+1.266K^{1/5}|x|-0.5 x^2 \label{eq:s_ti},
\end{align}
and the surface density for $x > x_m$ is given by equation~(\ref{eq:s_kepler}).
Since there is no torque density for $x>1.3$ in this case, the minimum surface density $\smin$ is given by $s(x=1.3)$.
Thus, we give $\smin$ as
\begin{align}
	\smin&=\exp\left( -0.854K^{2/5}+1.645K^{1/5}-0.845 \right) \label{eq:smin_tanigawa}.
\end{align}
Note that if $x_m < 1.3$, the whole region of the gap is the Keplerian rotating part and $\smin$ is given by equation~(\ref{eq:s_kepler}) with $x=1.3$.
It should be also noticed that the angular momentum conservation is not satisfied at the flat bottom region with $s=\smin$ of equation~(\ref{eq:smin_tanigawa}), as explained in the subsection~\ref{sub:effect_rayleigh} (and see also, Appendix~\ref{sec:kep_propagation}).
%We do not take care the angular momentum conservation in the derivation of equation~(\ref{eq:smin_tanigawa}).
%Indeed, the angular momentum is not conserved in the flat-bottom region (i.e.,$|x| < 1.3$).
%In the flat-bottom region, the surface densities given by equation~(\ref{eq:smin_tanigawa}) is inconsistent with equation~(\ref{eq:skep}) which is derived by the angular momentum conservation of equation~(\ref{eq:steady_disk}).
%Hence, the angular momentum conservation is broken down in the flat bottom region.

\section{Gap model in a Keplerian rotating disc with wave propagation} \label{sec:kep_propagation}
%In this section, assuming the Keplerian disc rotation, we consider gap structures in the case with wave propagation.
Here, by assuming the Keplerian disc rotation, we derive gap solutions in the case with wave propagation.
Following \cite{Tanigawa_Ikoma2007}, we consider the marginal condition when the Rayleigh condition is violated.
The angular momentum deposition rate is given by equation~(\ref{eq:nond_lambda_propagation}).
Ignoring deviation in $d\Omega/dR$ form the Keplerian and differentiating, we give
\begin{align}
	\dpar{s}{x}&=
	\begin{cases}
		\frac{\ttilde}{3}\frac{\hpp}{\wdep} & \for \ \xd-\frac{\wdep}{2\hpp}<|x|<\xd+\frac{\wdep}{2\hpp},\\
		0 & \mbox{otherwise},
	\end{cases}
	\label{eq:grads_kepler_propagation}
\end{align}
Integrating equation~(\ref{eq:grads_kepler_propagation}), we obtain the surface density without the Rayleigh condition as
\begin{align}
	s&=
	\begin{cases}
		1 & \for \ |x|>\xd+\frac{\wdep}{2\hpp},\\
		1-\frac{\ttilde \hpp}{3\wdep} \left[ \left( \xd+\frac{ \wdep}{2\hpp} \right) -  x \right]& \for \ \xd-\frac{\wdep}{2\hpp}<|x|<\xd+\frac{\wdep}{2\hpp},\\
		1-\frac{\ttilde}{3} &\for \ |x|<\xd-\frac{\wdep}{2\hpp}.
	\end{cases}
	\label{eq:skep_propagation}
\end{align}

Equation~(\ref{eq:skep_propagation}) does not satisfy the Rayleigh condition, especially for a large $K$ (or a large $\ttilde$).
First, the Rayleigh condition is violated near $|x|=\xd+\wdep/2\hpp$ because $ds/dx$ of equation~(\ref{eq:grads_kepler_propagation}) is not continueous there.
In order to make $ds/dx$ continueous, the marginally stable condition (eq.~[\ref{eq:gap_breaking_rayleigh}]) is used instead of equation~(\ref{eq:grads_kepler_propagation}) in the region where $\xd+\wdep-(\ttilde \hpp)/(3\wdep) < |x|< \xd+\wdep/2\hpp$.
In addition, the marginally stable condition should also be used near $|x|=\xd-\wdep/2\hpp$ for a large $\ttilde$.
From equation~(\ref{eq:grads_kepler_propagation}), we find that the Rayleigh condition is violated from $|x|=\xd-\wdep/2\hpp$ to $\xd+(\wdep/2\hpp)(1-6/\ttilde)+1$.
These two Rayleigh unstable regions are merged for $\ttilde > 3(\wdep/\hpp)$.
In such large-$K$ cases, since the marginally stable condition is used in the whole region of the angular momentum deposition, the minimum surface density is given by 
\begin{align}
	%\smin &= \exp\left[ -\frac{\hpp}{2\wdep} \left( \frac{\wdep}{\hpp}+2 \right)\right] \label{eq:smin_kepler_propagation_wrc}.
	\smin &= \exp\left[ -\left( \frac{1}{2}+\frac{\hpp}{\wdep} \right)\right] \label{eq:smin_kepler_propagation_wrc}.
\end{align}
which is independent of $K$.
This unrealistic minimum surface density does not satisfy equation~(\ref{eq:skep}) which is originated by the angular momentum conservation.
As indicated by \cite{Fung_etal2013}, the planetary torque should balance with the viscous angular momentum flux outside the gap for the angular momentum conservation.
In our formulation, two terms in the right-hand side of equation~(\ref{eq:sdif_zero_order2}) balance with each other in the bottom region (the left-hand side is negligibly small).
However, the minimum surface density given by equation~(\ref{eq:smin_kepler_propagation_wrc}) independent of $K$ breaks down such a balance.
Hence, equation~(\ref{eq:smin_kepler_propagation_wrc}) also violates the angular momentum conservation.

For a small $K$, equation~(\ref{eq:skep_propagation}) is approximately valid because the Rayleigh unstable region near $|x|=\xd + \wdep/2\hpp$ does not significantly affect $\smin$.
Substituting equation~(\ref{eq:skep_propagation}) into equation~(\ref{eq:relate_ttilde_K}), we give $\ttilde$ as
\begin{align}
	\ttilde & = 3\left[ \frac{9\Delta^3}{KC}\left( 1-\frac{KC\xd}{9\left[ \xd^2-(\wdep/2\hpp)^2 \right]^2} \right) +1 \right]^{-1}, \label{eq:ttilde_kepler_propagation}
\end{align}
Substituting equation~(\ref{eq:skep_propagation}) into equation~(\ref{eq:ttilde_kepler_propagation}),  we obtain $\smin$ for a small $K$ as
\begin{align}
	\smin&=\left[ \frac{CK}{9\Delta^3}\left( 1-\frac{KC\xd}{9\left[ \xd^2-(\wdep/2\hpp)^2 \right]^2} \right)^{-1}  + 1 \right]^{-1}.
	\label{eq:smin_kepler_propagation}
\end{align}

\section{Solutions for the linearized equation of gaps} \label{sec:particular_solution}
Here, we consider solutions to the following differential equation:
\begin{align}
	\ddpar{y}{x}-3y&=g(x) \label{eq:dequation1},
\end{align}
where $g(x)$ is an arbitrary odd function of $x$.
A general solution of this equation is given by a combination of homogeneous solutions, $e^{\pm \sqrt{3}x}$, and a particular solution.
We seek a solution which satisfies the boundary conditions of $y=0$ at $x=\pm \infty$.
Since $g(x)$ is odd in this case, the solution is an even function of $x$.
For simplicity, we consider the solution for $x>0$.
Because of the symmetry of the equation, the solution of $x<0$ can be obtained by inverting the sign of $x$ in the solution for $x>0$.

A particular solution $y_p(x)$ of equation~(\ref{eq:dequation1}) can be given by
\begin{align}
	y_p(x) = \frac{1}{2\sqrt{3}} \left[ e^{\sqrt{3}x} \int^{\infty}_{x} \right. & g(x') e^{-\sqrt{3}x'} dx' \nonumber \\
	& \left.- e^{-\sqrt{3}x} \int^{\infty}_{x} g(x') e^{\sqrt{3}x'} dx' \right] \label{eq:yp1}.
\end{align}
In the case with instantaneous wave damping (see eq.~[\ref{eq:linear_eq}]), $g(x)$ is given by
\begin{align}
	g(x)&=-
	\begin{cases}
		{\displaystyle \frac{C}{3|x|^3}}    &\for \ \ \ |x|>\Delta,\medspace \\
		{\displaystyle \frac{C}{3\Delta^3}} &\mbox{otherwise}.
	\end{cases}
	\label{eq:gx_insitu}
\end{align}
Substituting equation~(\ref{eq:gx_insitu}) into equation~(\ref{eq:yp1}), the particular solution for $x>\Delta$ is obtained as
\begin{align}
	y_{p}(x)&=\frac{\sqrt{3} C}{18}\left[ -\frac{\sqrt{3}}{x}+\frac{3}{2} \left\{ e^{-\sqrt{3}x}Ei(\sqrt{3}x)-e^{\sqrt{3}x}Ei(-\sqrt{3}x) \right\} \right] \label{eq:yp_outer},
\end{align}
where $Ei(ax)$ denotes the exponential integral function \citep[e.g.,][]{Handbook_math}.
For $x\leq \Delta$, the particular solution is given by
\begin{align}
	y_{p}(x)&=a_{+} e^{\sqrt{3}x}+a_{-} e^{-\sqrt{3}x} + \frac{C}{9\Delta^3} \label{eq:yp_inner},
\end{align}
where $a_{+}$ and $a_{-}$ are defined by
\begin{align}
	a_{\pm}&=\frac{C}{6} \left[  \mp Ei(\mp \sqrt{3}\Delta) - \frac{1}{\Delta} \left( \frac{1}{3\Delta^2}+\frac{1}{2} \right) e^{\mp \sqrt{3} \Delta} \right] \label{eq:acoef_pm}.
\end{align}
Using this particular solution given by equations~(\ref{eq:yp_outer}) and (\ref{eq:yp_inner}), we can obtain the general solution of equation~(\ref{eq:linear_eq}).
Since this particular solution vanishes at $x\rightarrow \infty$, the coefficient of the homogeneous solution $e^{\sqrt{3}x}$ is zero.
The coefficient of $e^{-\sqrt{3}x}$, $B$, is obtained as
\begin{align}
	B&=\frac{1}{\sqrt{3}}\left.\dpar{y_p}{x}\right|_{x=0}.
	\label{eq:integ_const}
\end{align}

\label{lastpage}
\end{document}